\documentclass[%
 reprint,
 amsmath,amssymb,
 aps,prl
]{revtex4-1}

\usepackage{graphicx}
\usepackage{dcolumn}
\usepackage{bm}
\usepackage{ulem}

\begin{document}


\title{A Search for Axionic Dark Matter Using the Magnetar PSR J1745$-$2900}

\author{Jeremy Darling}
 \affiliation{Center for Astrophysics and Space Astronomy \\
Department of Astrophysical and Planetary Sciences \\
University of Colorado, 389 UCB \\
Boulder, CO 80309-0389, USA}
 \email{jeremy.darling@colorado.edu}

\date{\today}

\begin{abstract}
We report on a search for dark matter axion conversion photons from the magnetosphere of the
Galactic Center magnetar PSR J1745$-$2900 using spectra obtained from the
Karl G. Jansky Very Large Array.\footnote{The National Radio Astronomy Observatory is a facility of the National Science Foundation operated under cooperative agreement by Associated Universities, Inc.}
No significant spectral features are detected.
  Using a hybrid model for PSR J1745$-$2900 and canonical assumptions
  about the dark matter density profile, we exclude axion models with axion-photon coupling
  $g_{a\gamma\gamma}> 6$--34~$\times 10^{-12}$~GeV$^{-1}$ with 95\%
  confidence over the mass ranges 4.2--8.4, 18.6--26.9, 33.0--41.4, 53.7--62.1, and 126.0--159.3~$\mu$eV.
  If there is a dark matter cusp, the limits reduce to $g_{a\gamma\gamma} > 6$--$34 \times10^{-14}$~GeV$^{-1}$,
  which overlap some axion models for the observed mass ranges $> 33$~$\mu$eV.
  These limits may be improved by modeling the stimulated emission that can boost the axion-photon conversion process.
\end{abstract}

\maketitle


\section{\label{sec:intro}Introduction}

The axion is a spin zero chargeless massive particle
introduced to address the strong $CP$ (charge-parity) problem in quantum chromodynamics
(QCD; \cite{Peccei1977,weinberg1978,wilczek1978}).  If they exist, QCD axions are likely to be produced in the early universe, are 
a promising cold dark matter candidate \cite{preskill1983,dine1983,abbott1983}, and may explain the cosmic matter-antimatter asymmetry
\cite{Co2020}.

Axions couple to QCD and electromagnetism.
The  electromagnetic coupling
$\mathcal{L}_{a\gamma\gamma} = -(1/4) g_{a\gamma\gamma} a\, F_{\mu\nu}\tilde{F}^{\mu\nu} =  g_{a\gamma\gamma} a\, {\mathbf E}\cdot{\mathbf B}$
suggests that axion-photon conversion can occur in the 
presence of magnetic fields, but the axion-photon coupling is weak ($g_{a\gamma\gamma}\sim 10^{-16}$~GeV$^{-1}$
for axion mass $m_a = 1$~$\mu$eV) \cite{Kim1979,Shifman1980,Dine1981,Zhitnitsky1980,sikivie1983}.
The relationship between $g_{a\gamma\gamma}$ and $m_a$ is linear, 
but the mass is not constrained; axion searches must span decades in $m_a$.

Recent axion searches include CAST, which searched for Solar axions \cite{arik2014,arik2015}, and haloscopes such as ADMX and HAYSTAC
that use narrow-band resonant cavities to detect dark matter axions \cite{asztalos2001,asztalos2010,brubaker2017,Zhong2018}.
There are also natural settings where telescopes may conduct sensitive and wide-band QCD axion searches
\cite{Arvanitaki2015,hook2018,huang2018,mukherjee2018,day2019,battye2020,leroy2020,edwards2020,mukherjee2020}.

The Galactic Center magnetar PSR J1745$-$2900 offers an ideal setting
for an axion conversion line search: it has a strong magnetic field ($1.6\times10^{14}$~G) \cite{mori2013}
and should see the highest possible dark matter flux \cite{hook2018}.
%
%
Axions will encounter a
plasma frequency at some radius that equals
its mass, and the axion can resonantly convert into a photon at that location \cite{hook2018}.
The most promising axion mass range, 
1--100~$\mu$eV, corresponds to radio frequencies 200 MHz to 20 GHz.

In this {\it Letter}, we present broad-band radio telescope observations of  PSR~J1745$-$2900.
We obtain 95\% confidence limits on resonant axion-photon conversion emission line flux density from the magnetar
spanning 40\% of the 1--38.5~GHz band.  We 
translate these flux limits into limits on $g_{a\gamma\gamma}$ versus $m_a$ based on a hybrid neutron star magnetosphere model and two
limiting-case 
Galactic dark matter profiles.  Finally, we discuss model caveats,  limitations of the observations, and future
observational and theoretical work to expand the $g_{a\gamma\gamma}$ vs.\ $m_a$ space probed by this technique.

\begin{table*}
  \caption{\label{tab:obs}Very Large Array Summary of Observations.}
\begin{ruledtabular}
\begin{tabular}{ccccccccrcl}
  Band & Frequency & Program & \multicolumn{2}{c}{Channel Width} & Median & Integration & Median Beam & PA & MJD\footnote{Modified Julian Date} & rms\footnote{The rms noise in spectra Gaussian-smoothed to channels of width $\Delta f$ (Column 5 and Equation \ref{eqn:df}).}\\
  & & & Obs. & Smoothed & Velocity\\
  & (GHz) & & (MHz) & (MHz) & (km~s$^{-1}$) & (s) & (arcsec) & ($^\circ$) & &  (mJy) \\ \hline
  L & 1.008--2.032 & 14A-231 & 1  & 10  & 2000 & 10591 & \ $2.22\times1.10$\footnote{The quoted L-band beam is the continuum beam; the synthesized beam in the spectral cube is highly variable due to RFI and the large fractional bandwidth.} & 1 & 56749 & 0.33\footnote{The rms noise includes residual unmitigated RFI.}\\
  C & 4.487--6.511  & 14A-231 & 2 & 14 & 763 & 10591  & $0.61\times0.28$ & $-$3 & 56749 & 0.099\\
  X & 7.987--10.011  & 14A-231 & 2 & 18 & 600 & 19148 & $0.34\times0.17$ & $-$1 & 56718 &  0.026\footnote{The rms noise omits the central RFI feature and band edges.} \\
Ku & 12.988--15.012 & 14A-231 & 2 & 20 & 428 & 16156 & $0.24\times0.11$ & $-$1 & 56726 &   0.027\\
Ka & 30.476--32.524 & 14A-232 & 2 & 26 & 247 & 17053  & $0.101\times0.050$ & $-$3 & 56725 &  0.100\\
Ka & 32.476--34.524 & 14A-232 & 2 & 26 & 233 & 17053  & $0.094\times0.046$ & $-$3 & 56725 &  0.116\\
Ka & 34.476--36.524 & 14A-232 & 2 & 28 & 236 & 17053  & $0.089\times0.045$ & $-$4 & 56725 &  0.165\\
Ka & 36.476--38.524 & 14A-232 & 2 & 28 & 224 & 17053  & $0.084\times0.042$ & $-$3 & 56725 &  0.152\\
\end{tabular}
\end{ruledtabular}
\end{table*}

\section{\label{sec:obs}Observations}

Interferometric observations of Sgr A* were obtained from the National Science Foundation's
Karl G. Jansky Very Large Array (VLA \footnote{The National Radio Astronomy Observatory is a facility of the National Science Foundation operated under cooperative agreement by Associated Universities, Inc.}) data archive.  We selected sessions:
(1) to maximize on-target integration time, (2) with adequate angular resolution to separate PSR J1745$-$2900
from Sgr A* (1.7'' in both coordinates), (3) to maximize total bandwidth, and (4) to adequately sample the expected emission line bandwidth.
 VLA A-configuration observations of Sgr A*,
 14A-231 and 14A-232, meet these criteria, spanning 40\% of the 1--38.5~GHz range (Table \ref{tab:obs}).
The gaps in this range are not covered by any extant observations that meet the above criteria.  


Each session used the flux density calibrator  J1331+3030 (3C286) and the bandpass calibrator J1733$-$1304.
For L- and C-band, the bandpass calibrator was also used for complex gain calibration.
The other bands used J1744$-$3116 for complex gain calibration.
Observations of the gain calibrator were interleaved with Sgr A* integrations with 8--30 min cadence.  Each band was observed
for a full transit of Sgr A*.  
The correlator was configured to obtain four polarization products (only the two parallel-hand products were used to form Stokes-I spectra below),
in 8 to 64 adjacent 64-channel spectral windows with 128 MHz bandwidth each.  The spectral windows were divided into 2 or 4 slightly overlapping baseband groups.  The L-band configuration differed:  it used spectral windows of 64 MHz bandwith and 1 MHz channels.  All sampling was
8-bit except for the Ka-band session, which used 3-bit sampling.  Correlator dump times were 1--5 s.  

\section{\label{sec:reduc}Data Reduction}

All data reduction tasks were performed using CASA \footnote{McMullin, J. P., Waters, B., Schiebel, D., Young, W., \& Golap, K. 2007, Astronomical Data Analysis Software and Systems XVI (ASP Conf. Ser. 376), ed. R. A. Shaw, F. Hill, \& D. J. Bell (San Francisco, CA: ASP), 127.  See also
\protect\url{https://science.nrao.edu/facilities/vla/data-processing/pipeline} for the CASA calibration pipeline and \protect\url{https://science.nrao.edu/facilities/vla/data-processing} for VLA calibration and analysis.}.  
After data flagging, we applied the flux, delay, atmospheric transmission, complex bandpass, and complex gain calibration  to the target field.
We then performed in-beam phase self-calibration using the Sgr A* continuum, which ranged from 0.5 to 5.6 Jy (L- to Ka-band).  We imaged the
continuum field and identified the continuum emission of PSR J1745$-$2900 (20--5 mJy in L to Ka bands) to confirm its coordinate offset from
Sgr A*.  

We fit a 2D Gaussian to the Sgr A* continuum to set the origin for relative astrometry to locate PSR J1745$-$2900 based on the bootstrap proper motion solution obtained by \cite{Bower2015} that assumes no acceleration or core shift.  At the observed angular resolution, Sgr A* is consistent with
being a point source.  Coordinate offsets from Sgr A* were $(\Delta\alpha,\Delta\delta) = (1.701,-1.679)$ arcsec in all epochs; offsets between epochs differed by no more than 1 mas.  The predicted offsets of PSR J1745$-$2900 from Sgr~A* are consistent with the observed continuum position of the magnetar.

We fit and subtracted a linear continuum spectrum
from the data in UV (image Fourier) space and formed spectral image cubes using pixels that over-sampled the
synthesized beams (which vary across each cube).
Sgr~A* shows narrow-band spectral structure after the continuum subtraction, some of which are attributable to known molecular
lines such as the 14.4881 GHz H$_2$CO and 36.795 GHz CH$_3$CN.
Sidelobes from these features are cleaned during cube deconvolution and do not significantly contaminate the magnetar spectrum.
The zero-mean magnetar spectra typically reach the theoretical noise.  

At the position of the magnetar, we extract a spectrum by summing pixels over the (frequency-dependent) synthesized 2D Gaussian beam.
We correct the spectrum for the beam size to capture the total flux density of a point source separately for every channel.  
In order to assess the significance of features in the magnetar spectrum, which can be highly variable channel-to-channel
due to radio frequency interference (RFI) or band edges, we create a noise spectrum by measuring the off-source rms sky noise in each
channel.  This allows us to create a significance spectrum for each band.  We scale the noise spectrum to match the spectral noise
measured toward PSR J1745$-$2900 (this correction is typically a few percent).

The fractional bandwidth predictions for the expected signal range from $\Delta f/f = (v_0/c)^2$  \cite{hook2018}, where $v_0$ is the axion velocity dispersion,
to  $(v_0/c)$ \cite{huang2018}.  We adopt the spinning mirror model whereby the Doppler broadening is dominated by the neutron star rotation \cite{battye2020},
producing, on average,
$\Delta f/f \simeq \Omega\, r_c \varepsilon^2/c$, where $\Omega$ is the neutron star's angular frequency, $\varepsilon$ is the eccentricity of the
elliptical critical surface (which we and \cite{battye2020} set to unity), and $r_c$ is the axion-photon conversion radius.  For the latter, we adopt the formulation of \cite{hook2018}, assuming polar
orientation angle $\theta = \pi/2$ and magnetic offset axis $\theta_m = 0$ (see below, where we marginalize over these unknown angles;
as \cite{leroy2020} show, ray-tracing is the technically correct approach, and tends to show less variation with these angles than the analytic treatment):
\begin{equation}
  r_c = 224~{\rm km}~\left(r_0 \over 10~{\rm km}\right) \left[ {B_0 \over 10^{14}~{\rm G}} {1\over 2\pi} {\Omega \over 1~{\rm Hz}} \left(4.1~\mu{\rm eV} \over m_a c^2\right)^2\right]^{1/3}
\end{equation}
where $r_0$ is the neutron star radius, $B_0$ is the magnetic field, and 4.1~$\mu$eV corresponds to an observed frequency of 1~GHz. 
This leads to an expected line bandwidth: 
\begin{equation}\label{eqn:df}
  \Delta f = 3.6~{\rm MHz} \left(\Omega \over 1~{\rm Hz}\right)^{4/3} \left(m_a c^2 \over 4.1~\mu{\rm eV}\right)^{1/3} \left(B_0 \over 10^{14}~{\rm G}\right)^{1/3}.
\end{equation}
The spin period of PSR J1745$-$2900 is 3.76~s \cite{kennea2013} and the magnetic field
is $1.6\times10^{14}$~G \cite{mori2013}.  The expected line width is therefore $\Delta f = 8.3~{\rm MHz} \times (m_a c^2 / 4.1~\mu{\rm eV})^{1/3}$.

The VLA correlator spectral windows have a strong response dropoff on edge channels and on the edges of the basebands.  These channels
lack the signal-to-noise for calibration and are flagged, which typically creates single-channel gaps in the spectrum and decreased
sensitivity on the spectrum edges.  The flagged channels are much narrower than the expected line width.  
We smooth the magnetar and noise spectra to the expected line width (Equation \ref{eqn:df}) using a Gaussian kernel \cite{astropy:2018}, and interpolate across missing channels during smoothing (the resulting statistics are not substantially altered by this loss of strict channel-to-channel independence).  In L-band there are gaps in the spectrum where RFI-flagged spectral regions are wider than the smoothing kernel.

Table \ref{tab:obs} lists synthesized beam parameters,  channel widths, and spectral rms noise values.  The Supplemental Material
\footnote{See Supplemental Material at [URL].} presents the magnetar flux, noise, and signal-to-noise spectra.

\begin{figure}[t]
\begin{centering}
  \includegraphics[scale=0.25,trim=20 20 20 0,clip=true]{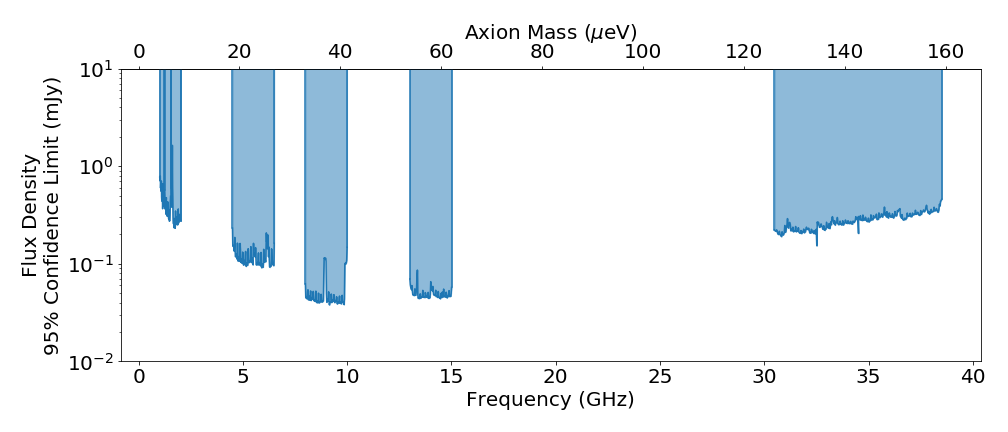}
\caption{
  95\% confidence limits on emission line flux density for channel width $\Delta f = 8.3~{\rm MHz} \times (m_a c^2 / 4.1~\mu{\rm eV})^{1/3}$.}
\label{fig:flux}
\end{centering}
\end{figure}

\section{\label{sec:results}Results}

We detect no significant single-channel emission features in the observed bands.
The only two channels above $3\sigma$ in emission were found in Ka-band (3.1$\sigma$ and 3.5$\sigma$).
These departures are consistent with Gaussian noise statistics.

After establishing the lack of significant detections in the observed spectra, we 
form  95\% confidence limits from the sky noise spectra. 
Figure \ref{fig:flux} presents these single-channel confidence limits.  These are model-independent 
limits on the flux density of photons produced by axion conversion in the magnetosphere of PSR J1745$-$2900
\footnote{Our assumption about the expected signal bandwidth does depend on models for the magnetar, but is $\mathcal{O}(v_0/c)$.}.
We present these limits to enable models not treated below to be translated into axion limits.

\section{\label{sec:analysis}Analysis}

Limits on the axion-photon coupling $g_{a\gamma\gamma}$ obtained from observed flux density limits depend on
the magnetar magnetospheric and axion-photon conversion model and on the behavior of the dark matter at the Galactic Center.
Our choices --- as well as alternatives --- are discussed below.

\subsection{\label{subsec:magnetar}The Magnetar Model}

Hook et al.\  \cite{hook2018} predict the photon flux from axion conversion in a neutron star magnetosphere using a variant of the
Goldreich and Julian \cite{goldreich1969} model.  The observed flux density depends on the local dark matter density
($\rho_\infty$) and its velocity dispersion ($v_0$) and on the neutron star mass, radius, magnetic field, rotation period, distance,
viewing angle with respect to the rotation axis, and magnetic misalignment angle from the spin axis
($M_{NS}$, $r_0$, $B_0$, $P$, $d$, $\theta$, and $\theta_m$, respectively).  It also depends on $m_a$ and
$g_{a\gamma\gamma}$. Equations 11--13 in \cite{hook2018}, modified for
the expected bandwidth (Equation \ref{eqn:df}), 
yield an expression for the emission line flux density normalized to fiducial values for a nearby pulsar:
%
%
\begin{widetext}
\begin{eqnarray}\label{eqn:flux}
  S_\nu = 1.2 \times10^{-6}\  {\rm mJy}\ \left(100\ {\rm pc} \over d\right)^2 
   \left(m_a \over 1\ {\rm GHz}\right)^{4/3} 
   \left(200\ {\rm km~s}^{-1}\over v_0\right) 
    \left(g_{a\gamma\gamma} \over 10^{-12}\ {\rm GeV}^{-1}\right)^2
    \left(r_0 \over 10\ {\rm km}\right)
     \nonumber \\
    \times \left(B_0 \over 10^{14}\ \rm{G}\right)^{1/3} \left(\Omega \over 1\ {\rm Hz}\right)^{-8/3}
    \left(\rho_\infty \over 0.3\ {\rm GeV~cm}^{-3}\right) \left(M_{NS} \over 1\ {\rm M}_\odot\right)
    {3 (\hat{m}\cdot\hat{r})^2 + 1 \over |3\cos\theta\ \hat{m}\cdot\hat{r}-\cos\theta_m|^{4/3}} {v_c\over c}
\end{eqnarray} 
\end{widetext}
where $\hat{m}\cdot\hat{r} = \cos\theta_m \cos\theta + \sin\theta_m \sin\theta \cos\Omega t$.
The axion velocity $v_c$ at the conversion point $r_c$ (see Supplement) 
is included to correct the conversion probability presented in \cite{hook2018}, as discussed in \cite{leroy2020}.
For a {\it given} line sensitivity (or observed flux density limit spectrum), one can obtain a track in $g_{a\gamma\gamma}$ versus $m_a$ space
above which $g_{a\gamma\gamma}$ is excluded at some confidence level.

For PSR J1745$-$2900,
we assume a radius $r_0 = 10$~km and mass $M_{NS} = 1$~M$_\odot$.
Using the Navarro-Frenk-White (NFW \cite{NFW}) Galactic dark matter profile $\gamma=1$ model of \cite{McMillan2017},  which has
Galactic Center distance $d = 8.2$~kpc (in agreement with the S2 stellar orbital measurement around Sgr A* \cite{Abuter2019}),
scale radius $r_s = 18.6$~kpc, and local dark matter energy density $\rho_\odot = 0.38$~GeV~cm$^{-3}$ (see below),
one obtains an axion mass-dependent expression for $g_{a\gamma\gamma}$:
\begin{widetext}
\begin{eqnarray}\label{eqn:g_agg} 
  g_{a\gamma\gamma} =  3 \times 10^{-11}\ {\rm GeV}^{-1}\
  \left(S_\nu \over 10\ \mu{\rm Jy}\right)^{1/2}\ 
  \left(m_a \over 1\ {\rm GHz}\right)^{-2/3} 
  \left(v_0\over 200\ {\rm km~s}^{-1}\right)^{1/2}\ \nonumber \\
  \times \left(\rho_\infty \over 6.5\times10^4\  {\rm GeV~cm}^{-3}\right)^{-1/2} \ 
  \left.\underbrace{\left({3 (\hat{m}\cdot\hat{r})^2 + 1 \over |3\cos\theta\ \hat{m}\cdot\hat{r}-\cos\theta_m|^{4/3}} {v_c\over c} \right)}_{(i)}\right.^{-1/2}.
\end{eqnarray}
\end{widetext}
The angular term $(i)$ in Equation \ref{eqn:g_agg} relies on the unknown viewing and magnetic field misalignment angles, so we
marginalize over all possible orientations while keeping track of the fraction of the magnetar spin period (if any) that has $r_c < r_0$
(see Supplement).
If the conversion radius $r_c$ is less than the neutron star radius, then
no conversion occurs and the photon flux vanishes \cite{hook2018}.  The analytic value of $r_c < r_0$ is a conservative estimate; ray-tracing suggests that
the actual axion-photon conversion can occur over a larger range of angles \cite{leroy2020}.  Contrary to \cite{hook2018}, we find that conversion can still
occur up to and beyond 10~GHz.  At the low end of the observed Ka-band range, 2.1\% of all possible $(\theta,\theta_m)$ have $r_c < r_0$
at all times; at the high end of the band, this increases to 5.6\%.  At the high end of the observed Ku band, the fraction of orientations with no emission
decreases to 0.16\%.

\subsection{\label{subsec:dm}Axion Constraints from Dark Matter Models}

Following \cite{hook2018}, we use two dark matter models --- a generic NFW model
and the NFW model plus a central dark matter spike --- that roughly bound the weakest and strongest constraints on
$g_{a\gamma\gamma}$ given the magnetar model unless the Galactic Center dark matter is cored (see below and Discussion).  
We use the NFW dark matter model from \cite{McMillan2017} with $\gamma=1$, $R_0 = 8.2$~kpc,
$r_s = 18.6$~kpc, $v_0 = 300$~km~s$^{-1}$, and $\rho_\odot = 0.38$~GeV~cm$^{-3}$, which predicts a dark matter density of
$6.5\times10^4$~GeV~cm$^{-3}$ at the 0.1~pc radial distance of the magnetar.  
The assumed physical distance of PSR J1745$-$2900 from Sgr A* --- itself assumed to reside at the center of the dark matter distribution --- is the projected distance (absent an acceleration measurement, one cannot determine the physical distance between Sgr A* and the magnetar \cite{Bower2015},
but if the magnetar is gravitationally bound to Sgr A*, then the projected separation is likely to be a good approximation for the physical separation).  
For the dark matter spike,
we use the above NFW profile modified to include a spike radial index $\gamma_{sp} = 7/3$ and extent $R_{sp}=0.1$~kpc, which
corresponds to a 99.7\% upper limit on deviations from a black hole-only orbit of the S2 star about Sgr A*  \cite{Lacroix2018}.  This model predicts a
dark matter density of $6.4\times10^8$~GeV~cm$^{-3}$ at the magnetar position.  All else equal, this dark
matter spike model can be considered to be a best-case constraint on $g_{a\gamma\gamma}$.  The two models span a factor of $10^4$
in dark matter density, which is a factor of 100 in the $g_{a\gamma\gamma}$ constraint.  Future studies of the dynamics
in the central parsec should improve constraints on the dark matter density encountered by PSR~J1745$-$2900.

\begin{figure*}
\begin{centering}
  \includegraphics[scale=0.5,trim=20 20 20 0,clip=true]{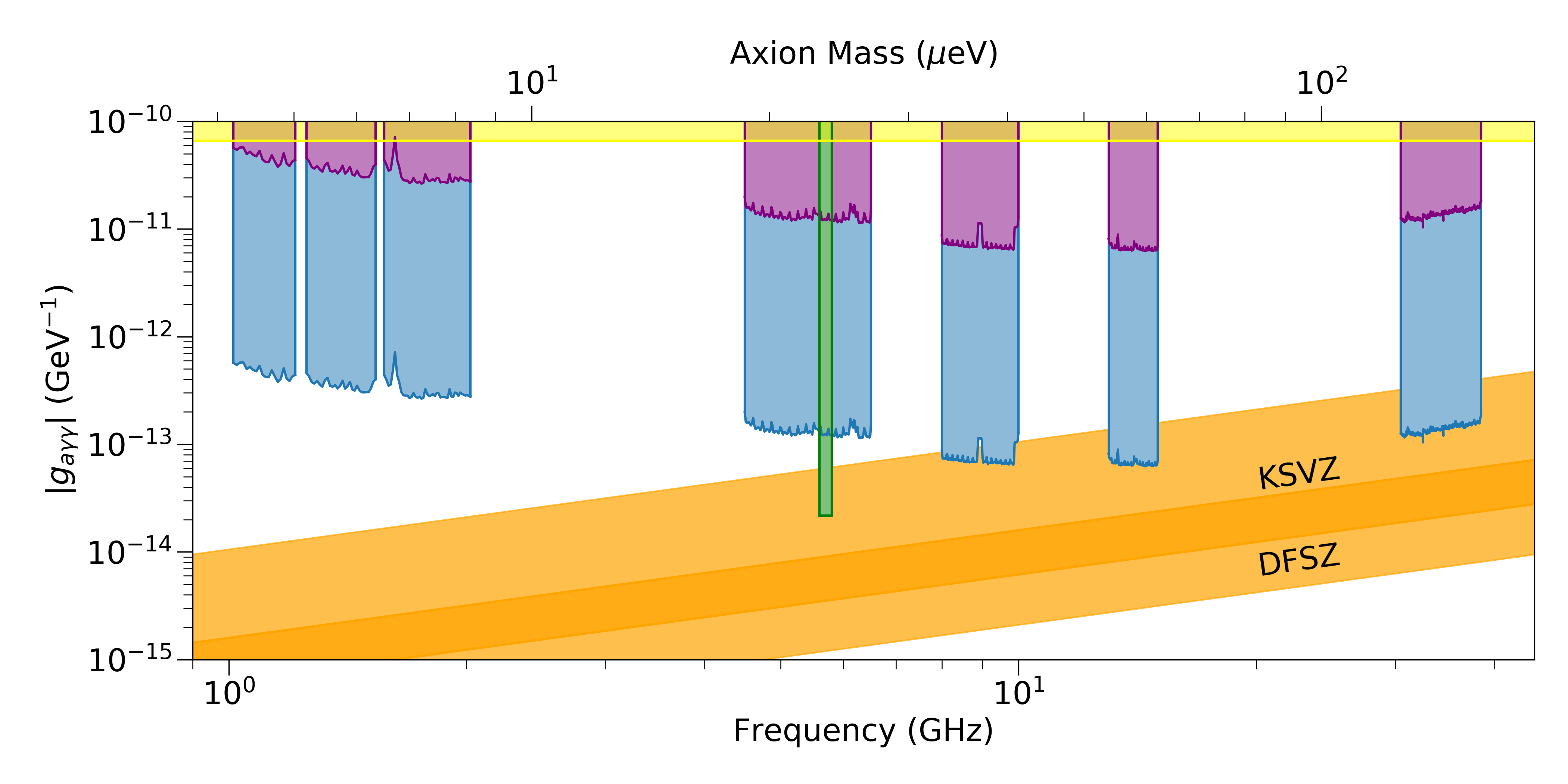}
\caption{
 95\% confidence limits on $g_{a\gamma\gamma}$ for a generic NFW $\gamma=1$ model with
 $\rho_\odot = 0.38$~GeV~cm$^{-3}$ 
 (purple, upper) and the same NFW model plus a central 100~pc dark matter spike (blue, lower).
 The green bar shows the HAYSTAC limit \cite{Zhong2018}, which has been scaled from a local axion density of 0.45~GeV~cm$^{-3}$ to 0.38~GeV~cm$^{-3}$,
 and the yellow bar shows the CAST 95\% confidence limit \cite{Anastassopoulos2017}.
 Orange loci indicate a range of possible QCD axion models \cite{Luzio2017}, including the canonical KSVZ and DFSZ models
  \cite{Kim1979,Shifman1980,Dine1981,Zhitnitsky1980}.  
}
\label{fig:limits}
\end{centering}
\end{figure*}

Table \ref{tab:g_agg} lists the median 95\% confidence limits on $|g_{a\gamma\gamma}|$ obtained from these end-case dark matter models
for each observed band.  Figure \ref{fig:limits} shows the limits for the two models versus frequency and axion mass.  The limits obtained for
the standard NFW profile reach $|g_{a\gamma\gamma}| \simeq 6\times10^{-12}$~GeV$^{-1}$, which is a factor of $\sim$40 above the
strongest-coupling theoretical model \cite{Luzio2017}.  The limits obtained for
the dark matter spike model impinge on the family of theoretical models \cite{Luzio2017} for $m_a$ in 33.0--41.4~$\mu$eV,
53.7--62.1~$\mu$eV,  and 126.0--159.3~$\mu$eV, but do not exclude the canonical KSVZ or DFSZ models
\cite{Kim1979,Shifman1980,Dine1981,Zhitnitsky1980}.  If the Galactic dark matter has a flat profile inward of 0.5 kpc (i.e. it is ``cored''), then
the predicted dark matter density is 12~GeV~cm$^{-3}$ at the magnetar,
the limits on $g_{a\gamma\gamma}$ are a factor of 100 above the NFW profile limits, and the limits are less constraining than the CAST limits.

%
%
%
%

\begin{table}
  \caption{\label{tab:g_agg} Limits on $g_{a\gamma\gamma}$ for two dark matter profile models.  }
\begin{ruledtabular}
\begin{tabular}{ccc}
  Axion Mass  & \multicolumn{2}{c}{Median $|g_{a\gamma\gamma}|$ 95\% Confidence Limits} \\[2pt] \cline{2-3} \\[-8pt] 
      ($\mu$eV)  & NFW Profile & DM Spike \\
  & (GeV$^{-1}$) & (GeV$^{-1}$) \\[2pt] \hline \\[-8pt]
   4.2--8.4\footnote{There are gaps in the coverage of this mass range (see Figure \ref{fig:limits}).} & $3.4 \times 10^{-11}$ &  $3.4 \times 10^{-13}$ \\
  18.6--26.9 &  $1.3 \times 10^{-11}$ &   $1.3 \times 10^{-13}$\\
  33.0--41.4 &  $6.9 \times 10^{-12}$ &   $7.0 \times 10^{-14}$\\
  53.7--62.1 &  $6.5 \times 10^{-12}$ &   $6.5 \times 10^{-14}$\\
126.0--159.3 & $1.4 \times 10^{-11}$ &   $1.4 \times 10^{-13}$\\
\end{tabular}
\end{ruledtabular}
\end{table}

\section{\label{sec:discussion}Discussion}

The limits on $g_{a\gamma\gamma}$ presented here are less stringent than those predicted by \cite{hook2018}, even
after scaling the telescope sensitivity and integration time.  This is due to the expected bandwidth of the conversion
emission line; we used a line width roughly $10^3$--$10^4$ times larger, following the rotating mirror conversion region treatment of
\cite{battye2020}, which leads to a factor of $\sim$300 in $g_{a\gamma\gamma}$.  This bandwidth is likely the most conservative choice,
and a more physical model for the axion-photon conversion bandwidth will require detailed ray-tracing and a more sophisticated
magnetosphere model \cite{leroy2020}. 

The magnetospheres of magnetars must be more complicated than current models assume.  Moreover, the resonant conversion
of axions into photons will likely be boosted by stimulated emission caused by the ambient photon bath \cite{Caputo2019}.
The expected conversion emission line would be enhanced by the photon occupation number $f_\gamma = \pi^2 \rho_\gamma / E_\gamma^3$, where
$\rho_\gamma$ is the specific energy density of stimulating photons (which must match the
conversion photons in energy and momentum).   Including the contribution from stimulated emission, which is expected to be
particularly important in the Galactic Center environment \cite{Caputo2019}, will significantly improve
the constraints on $g_{a\gamma\gamma}$ without the need for additional observations.  Numerical simulation may be required
to correctly characterize the role of stimulated emission in resonant axion-photon conversion in neutron star magnetospheres.

Finally, the constraints on $g_{a\gamma\gamma}$ depend critically on the dark matter energy density in the inner parsec, which remains
poorly constrained compared to the other parameters in Equations \ref{eqn:flux} and \ref{eqn:g_agg} (which have uncertainties of order
10--50\%).  Cuspy versus cored dark matter profiles depend on the dominant baryon physics.
Baryons can contract and steepen the central dark matter density, and there is evidence that this has happened in the Galaxy \cite{cautun2020},
but baryon feedback (e.g.\ star formation and supernovae) can flatten the central dark matter profile.
While a multi-kpc core is disfavored \cite{hooper2017},
current studies cannot yet probe the inner kpc, so the dark matter density at the magnetar remains an extrapolation.  
It seems likely, however, that observations of stars and gas combined with
dynamical models will soon provide meaningful measurements of the role of dark matter in the Galactic Center.
The sensitivity of future telescopic observations will also improve with the advent of large
collecting-area facilities such as the Square Kilometer Array, although added sensitivity above a few GHz would require a next-generation VLA.

\section{\label{sec:conclusions}Conclusions}

We have demonstrated how radio telescope spectral observations of the magnetar PSR~J1745$-$2900 can produce broad-band limits on
axion-photon coupling.  This work can be expanded by observing more of the 1--50~GHz spectral window and obtaining
longer integration times.  Future observations could also increase signal-to-noise by observing in a spectral line pulsar mode, which
requires fine time sampling of the neutron star spin period.

The VLA interferometer in its most extended configuration spatially filters out any extended axion conversion spectral line produced in the 
collective magnetospheres of the as-yet undetected population of Galactic Center neutron stars.  An optimal
strategy to detect this extended signal would depend on the areal extent of the emission, the magnetized neutron
star population, and the dark matter density in the Galactic Center \cite{Safdi2019}.  

There are parts of the radio spectrum that will never be observable for axion signals from the ground, such as
global positioning bands, large parts of the sub-GHz spectrum due to RFI, the 22 GHz atmospheric water band,
and the molecular oxygen band spanning 50--70~GHz.  These regions merit special scrutiny by laboratory
experiments.  Magnetar observations are further limited by the requirement that axions cannot be converted
inside the neutron star, above roughly 50~GHz.

\begin{acknowledgments}
  We thank the observing, archive, and computing staff at the NRAO who made this work possible.  We also thank
  Konrad Lehnert, Marco Chianese, Andrea Caputo, and Richard Battye for helpful discussions.
  This research made use of CASA \cite{CASA}, NumPy \cite{NumPy}, Matplotlib \cite{Matplotlib}, and
  Astropy \footnote{http://www.astropy.org}, a community-developed core Python package for Astronomy \citep{astropy:2013, astropy:2018}.
\end{acknowledgments}



\bibliography{ms}

\providecommand{\noopsort}[1]{}\providecommand{\singleletter}[1]{#1}%
\begin{thebibliography}{52}%
\makeatletter
\providecommand \@ifxundefined [1]{%
 \@ifx{#1\undefined}
}%
\providecommand \@ifnum [1]{%
 \ifnum #1\expandafter \@firstoftwo
 \else \expandafter \@secondoftwo
 \fi
}%
\providecommand \@ifx [1]{%
 \ifx #1\expandafter \@firstoftwo
 \else \expandafter \@secondoftwo
 \fi
}%
\providecommand \natexlab [1]{#1}%
\providecommand \enquote  [1]{``#1''}%
\providecommand \bibnamefont  [1]{#1}%
\providecommand \bibfnamefont [1]{#1}%
\providecommand \citenamefont [1]{#1}%
\providecommand \href@noop [0]{\@secondoftwo}%
\providecommand \href [0]{\begingroup \@sanitize@url \@href}%
\providecommand \@href[1]{\@@startlink{#1}\@@href}%
\providecommand \@@href[1]{\endgroup#1\@@endlink}%
\providecommand \@sanitize@url [0]{\catcode `\\12\catcode `\$12\catcode
  `\&12\catcode `\#12\catcode `\^12\catcode `\_12\catcode `\%12\relax}%
\providecommand \@@startlink[1]{}%
\providecommand \@@endlink[0]{}%
\providecommand \url  [0]{\begingroup\@sanitize@url \@url }%
\providecommand \@url [1]{\endgroup\@href {#1}{\urlprefix }}%
\providecommand \urlprefix  [0]{URL }%
\providecommand \Eprint [0]{\href }%
\providecommand \doibase [0]{http://dx.doi.org/}%
\providecommand \selectlanguage [0]{\@gobble}%
\providecommand \bibinfo  [0]{\@secondoftwo}%
\providecommand \bibfield  [0]{\@secondoftwo}%
\providecommand \translation [1]{[#1]}%
\providecommand \BibitemOpen [0]{}%
\providecommand \bibitemStop [0]{}%
\providecommand \bibitemNoStop [0]{.\EOS\space}%
\providecommand \EOS [0]{\spacefactor3000\relax}%
\providecommand \BibitemShut  [1]{\csname bibitem#1\endcsname}%
\let\auto@bib@innerbib\@empty
\bibitem [{\citenamefont {{Peccei}}\ and\ \citenamefont
  {{Quinn}}(1977)}]{Peccei1977}%
  \BibitemOpen
  \bibfield  {author} {\bibinfo {author} {\bibfnamefont {R.~D.}\ \bibnamefont
  {{Peccei}}}\ and\ \bibinfo {author} {\bibfnamefont {H.~R.}\ \bibnamefont
  {{Quinn}}},\ }\href {\doibase 10.1103/PhysRevLett.38.1440} {\bibfield
  {journal} {\bibinfo  {journal} {\prl}\ }\textbf {\bibinfo {volume} {38}},\
  \bibinfo {pages} {1440} (\bibinfo {year} {1977})}\BibitemShut {NoStop}%
\bibitem [{\citenamefont {Weinberg}(1978)}]{weinberg1978}%
  \BibitemOpen
  \bibfield  {author} {\bibinfo {author} {\bibfnamefont {S.}~\bibnamefont
  {Weinberg}},\ }\href@noop {} {\bibfield  {journal} {\bibinfo  {journal}
  {Phys.\ Rev.\ Lett.}\ }\textbf {\bibinfo {volume} {40}},\ \bibinfo {pages}
  {223} (\bibinfo {year} {1978})}\BibitemShut {NoStop}%
\bibitem [{\citenamefont {Wilczek}(1978)}]{wilczek1978}%
  \BibitemOpen
  \bibfield  {author} {\bibinfo {author} {\bibfnamefont {F.}~\bibnamefont
  {Wilczek}},\ }\href@noop {} {\bibfield  {journal} {\bibinfo  {journal}
  {Phys.\ Rev.\ Lett.}\ }\textbf {\bibinfo {volume} {40}},\ \bibinfo {pages}
  {279} (\bibinfo {year} {1978})}\BibitemShut {NoStop}%
\bibitem [{\citenamefont {Preskill}\ \emph {et~al.}(1983)\citenamefont
  {Preskill}, \citenamefont {Wise},\ and\ \citenamefont
  {Wilczek}}]{preskill1983}%
  \BibitemOpen
  \bibfield  {author} {\bibinfo {author} {\bibfnamefont {J.}~\bibnamefont
  {Preskill}}, \bibinfo {author} {\bibfnamefont {M.~B.}\ \bibnamefont {Wise}},
  \ and\ \bibinfo {author} {\bibfnamefont {F.}~\bibnamefont {Wilczek}},\
  }\href@noop {} {\bibfield  {journal} {\bibinfo  {journal} {Phys.\ Lett.\ B}\
  }\textbf {\bibinfo {volume} {120}},\ \bibinfo {pages} {127} (\bibinfo {year}
  {1983})}\BibitemShut {NoStop}%
\bibitem [{\citenamefont {Dine}\ and\ \citenamefont
  {Fischler}(1983)}]{dine1983}%
  \BibitemOpen
  \bibfield  {author} {\bibinfo {author} {\bibfnamefont {M.}~\bibnamefont
  {Dine}}\ and\ \bibinfo {author} {\bibfnamefont {W.}~\bibnamefont
  {Fischler}},\ }\href {\doibase https://doi.org/10.1016/0370-2693(83)90639-1}
  {\bibfield  {journal} {\bibinfo  {journal} {Physics Letters B}\ }\textbf
  {\bibinfo {volume} {120}},\ \bibinfo {pages} {137 } (\bibinfo {year}
  {1983})}\BibitemShut {NoStop}%
\bibitem [{\citenamefont {{Abbott}}\ and\ \citenamefont
  {{Sikivie}}(1983)}]{abbott1983}%
  \BibitemOpen
  \bibfield  {author} {\bibinfo {author} {\bibfnamefont {L.~F.}\ \bibnamefont
  {{Abbott}}}\ and\ \bibinfo {author} {\bibfnamefont {P.}~\bibnamefont
  {{Sikivie}}},\ }\href {\doibase 10.1016/0370-2693(83)90638-X} {\bibfield
  {journal} {\bibinfo  {journal} {Physics Letters B}\ }\textbf {\bibinfo
  {volume} {120}},\ \bibinfo {pages} {133} (\bibinfo {year}
  {1983})}\BibitemShut {NoStop}%
\bibitem [{\citenamefont {{Co}}\ and\ \citenamefont
  {{Harigaya}}(2019)}]{Co2020}%
  \BibitemOpen
  \bibfield  {author} {\bibinfo {author} {\bibfnamefont {R.~T.}\ \bibnamefont
  {{Co}}}\ and\ \bibinfo {author} {\bibfnamefont {K.}~\bibnamefont
  {{Harigaya}}},\ }\href@noop {} {\bibfield  {journal} {\bibinfo  {journal}
  {arXiv e-prints}\ ,\ \bibinfo {eid} {arXiv:1910.02080}} (\bibinfo {year}
  {2019})},\ \Eprint {http://arxiv.org/abs/1910.02080} {arXiv:1910.02080
  [hep-ph]} \BibitemShut {NoStop}%
\bibitem [{\citenamefont {{Kim}}(1979)}]{Kim1979}%
  \BibitemOpen
  \bibfield  {author} {\bibinfo {author} {\bibfnamefont {J.~E.}\ \bibnamefont
  {{Kim}}},\ }\href {\doibase 10.1103/PhysRevLett.43.103} {\bibfield  {journal}
  {\bibinfo  {journal} {\prl}\ }\textbf {\bibinfo {volume} {43}},\ \bibinfo
  {pages} {103} (\bibinfo {year} {1979})}\BibitemShut {NoStop}%
\bibitem [{\citenamefont {{Shifman}}\ \emph {et~al.}(1980)\citenamefont
  {{Shifman}}, \citenamefont {{Vainshtein}},\ and\ \citenamefont
  {{Zakharov}}}]{Shifman1980}%
  \BibitemOpen
  \bibfield  {author} {\bibinfo {author} {\bibfnamefont {M.~A.}\ \bibnamefont
  {{Shifman}}}, \bibinfo {author} {\bibfnamefont {A.~I.}\ \bibnamefont
  {{Vainshtein}}}, \ and\ \bibinfo {author} {\bibfnamefont {V.~I.}\
  \bibnamefont {{Zakharov}}},\ }\href {\doibase 10.1016/0550-3213(80)90209-6}
  {\bibfield  {journal} {\bibinfo  {journal} {Nuclear Physics B}\ }\textbf
  {\bibinfo {volume} {166}},\ \bibinfo {pages} {493} (\bibinfo {year}
  {1980})}\BibitemShut {NoStop}%
\bibitem [{\citenamefont {{Dine}}\ \emph {et~al.}(1981)\citenamefont {{Dine}},
  \citenamefont {{Fischler}},\ and\ \citenamefont {{Srednicki}}}]{Dine1981}%
  \BibitemOpen
  \bibfield  {author} {\bibinfo {author} {\bibfnamefont {M.}~\bibnamefont
  {{Dine}}}, \bibinfo {author} {\bibfnamefont {W.}~\bibnamefont {{Fischler}}},
  \ and\ \bibinfo {author} {\bibfnamefont {M.}~\bibnamefont {{Srednicki}}},\
  }\href {\doibase 10.1016/0370-2693(81)90590-6} {\bibfield  {journal}
  {\bibinfo  {journal} {Physics Letters B}\ }\textbf {\bibinfo {volume}
  {104}},\ \bibinfo {pages} {199} (\bibinfo {year} {1981})}\BibitemShut
  {NoStop}%
\bibitem [{\citenamefont {Zhitnitskij}(1980)}]{Zhitnitsky1980}%
  \BibitemOpen
  \bibfield  {author} {\bibinfo {author} {\bibfnamefont {A.~R.}\ \bibnamefont
  {Zhitnitskij}},\ }\href
  {http://inis.iaea.org/search/search.aspx?orig{\_}q=RN:11563932} {\bibfield
  {journal} {\bibinfo  {journal} {Yadernaya Fizika}\ }\textbf {\bibinfo
  {volume} {31}},\ \bibinfo {pages} {497} (\bibinfo {year} {1980})}\BibitemShut
  {NoStop}%
\bibitem [{\citenamefont {{Sikivie}}(1983)}]{sikivie1983}%
  \BibitemOpen
  \bibfield  {author} {\bibinfo {author} {\bibfnamefont {P.}~\bibnamefont
  {{Sikivie}}},\ }\href {\doibase 10.1103/PhysRevLett.51.1415} {\bibfield
  {journal} {\bibinfo  {journal} {\prl}\ }\textbf {\bibinfo {volume} {51}},\
  \bibinfo {pages} {1415} (\bibinfo {year} {1983})}\BibitemShut {NoStop}%
\bibitem [{\citenamefont {Arik}\ and\ \citenamefont {et~al.
  (CAST)}(2014)}]{arik2014}%
  \BibitemOpen
  \bibfield  {author} {\bibinfo {author} {\bibfnamefont {M.}~\bibnamefont
  {Arik}}\ and\ \bibinfo {author} {\bibnamefont {et~al. (CAST)}},\ }\href@noop
  {} {\bibfield  {journal} {\bibinfo  {journal} {Phys.\ Rev.\ Lett.}\ }\textbf
  {\bibinfo {volume} {112}},\ \bibinfo {pages} {091302} (\bibinfo {year}
  {2014})}\BibitemShut {NoStop}%
\bibitem [{\citenamefont {Arik}\ and\ \citenamefont {et~al.
  (CAST)}(2015)}]{arik2015}%
  \BibitemOpen
  \bibfield  {author} {\bibinfo {author} {\bibfnamefont {M.}~\bibnamefont
  {Arik}}\ and\ \bibinfo {author} {\bibnamefont {et~al. (CAST)}},\ }\href@noop
  {} {\bibfield  {journal} {\bibinfo  {journal} {Phys.\ Rev.\ D}\ }\textbf
  {\bibinfo {volume} {92}},\ \bibinfo {pages} {021101} (\bibinfo {year}
  {2015})}\BibitemShut {NoStop}%
\bibitem [{\citenamefont {Asztalos}\ and\ \citenamefont {et~al.
  (ADMX)}(2001)}]{asztalos2001}%
  \BibitemOpen
  \bibfield  {author} {\bibinfo {author} {\bibfnamefont {S.~J.}\ \bibnamefont
  {Asztalos}}\ and\ \bibinfo {author} {\bibnamefont {et~al. (ADMX)}},\
  }\href@noop {} {\bibfield  {journal} {\bibinfo  {journal} {Phys.\ Rev.\ D}\
  }\textbf {\bibinfo {volume} {64}},\ \bibinfo {pages} {092003} (\bibinfo
  {year} {2001})}\BibitemShut {NoStop}%
\bibitem [{\citenamefont {Asztalos}\ and\ \citenamefont
  {et~al.}(2010)}]{asztalos2010}%
  \BibitemOpen
  \bibfield  {author} {\bibinfo {author} {\bibfnamefont {S.~J.}\ \bibnamefont
  {Asztalos}}\ and\ \bibinfo {author} {\bibnamefont {et~al.}},\ }\href@noop {}
  {\bibfield  {journal} {\bibinfo  {journal} {Phys.\ Rev.\ Lett.}\ }\textbf
  {\bibinfo {volume} {104}},\ \bibinfo {pages} {041301} (\bibinfo {year}
  {2010})}\BibitemShut {NoStop}%
\bibitem [{\citenamefont {Brubaker}\ and\ \citenamefont
  {et~al.}(2017)}]{brubaker2017}%
  \BibitemOpen
  \bibfield  {author} {\bibinfo {author} {\bibfnamefont {B.~M.}\ \bibnamefont
  {Brubaker}}\ and\ \bibinfo {author} {\bibnamefont {et~al.}},\ }\href@noop {}
  {\bibfield  {journal} {\bibinfo  {journal} {Phys.\ Rev.\ Lett.}\ }\textbf
  {\bibinfo {volume} {118}},\ \bibinfo {pages} {061302} (\bibinfo {year}
  {2017})}\BibitemShut {NoStop}%
\bibitem [{\citenamefont {Zhong}\ \emph {et~al.}(2018)\citenamefont {Zhong},
  \citenamefont {{Al Kenany}}, \citenamefont {Backes}, \citenamefont
  {Brubaker}, \citenamefont {Cahn}, \citenamefont {Carosi}, \citenamefont
  {Gurevich}, \citenamefont {Kindel}, \citenamefont {Lamoreaux}, \citenamefont
  {Lehnert}, \citenamefont {Lewis}, \citenamefont {Malnou}, \citenamefont
  {Maruyama}, \citenamefont {Palken}, \citenamefont {Rapidis}, \citenamefont
  {Root}, \citenamefont {Simanovskaia}, \citenamefont {Shokair}, \citenamefont
  {Speller}, \citenamefont {Urdinaran},\ and\ \citenamefont {{Van
  Bibber}}}]{Zhong2018}%
  \BibitemOpen
  \bibfield  {author} {\bibinfo {author} {\bibfnamefont {L.}~\bibnamefont
  {Zhong}}, \bibinfo {author} {\bibfnamefont {S.}~\bibnamefont {{Al Kenany}}},
  \bibinfo {author} {\bibfnamefont {K.~M.}\ \bibnamefont {Backes}}, \bibinfo
  {author} {\bibfnamefont {B.~M.}\ \bibnamefont {Brubaker}}, \bibinfo {author}
  {\bibfnamefont {S.~B.}\ \bibnamefont {Cahn}}, \bibinfo {author}
  {\bibfnamefont {G.}~\bibnamefont {Carosi}}, \bibinfo {author} {\bibfnamefont
  {Y.~V.}\ \bibnamefont {Gurevich}}, \bibinfo {author} {\bibfnamefont {W.~F.}\
  \bibnamefont {Kindel}}, \bibinfo {author} {\bibfnamefont {S.~K.}\
  \bibnamefont {Lamoreaux}}, \bibinfo {author} {\bibfnamefont {K.~W.}\
  \bibnamefont {Lehnert}}, \bibinfo {author} {\bibfnamefont {S.~M.}\
  \bibnamefont {Lewis}}, \bibinfo {author} {\bibfnamefont {M.}~\bibnamefont
  {Malnou}}, \bibinfo {author} {\bibfnamefont {R.~H.}\ \bibnamefont
  {Maruyama}}, \bibinfo {author} {\bibfnamefont {D.~A.}\ \bibnamefont
  {Palken}}, \bibinfo {author} {\bibfnamefont {N.~M.}\ \bibnamefont {Rapidis}},
  \bibinfo {author} {\bibfnamefont {J.~R.}\ \bibnamefont {Root}}, \bibinfo
  {author} {\bibfnamefont {M.}~\bibnamefont {Simanovskaia}}, \bibinfo {author}
  {\bibfnamefont {T.~M.}\ \bibnamefont {Shokair}}, \bibinfo {author}
  {\bibfnamefont {D.~H.}\ \bibnamefont {Speller}}, \bibinfo {author}
  {\bibfnamefont {I.}~\bibnamefont {Urdinaran}}, \ and\ \bibinfo {author}
  {\bibfnamefont {K.~A.}\ \bibnamefont {{Van Bibber}}},\ }\href {\doibase
  10.1103/PhysRevD.97.092001} {\bibfield  {journal} {\bibinfo  {journal}
  {Physical Review D}\ }\textbf {\bibinfo {volume} {97}},\ \bibinfo {pages}
  {092001} (\bibinfo {year} {2018})}\BibitemShut {NoStop}%
\bibitem [{\citenamefont {{Arvanitaki}}\ \emph {et~al.}(2015)\citenamefont
  {{Arvanitaki}}, \citenamefont {{Baryakhtar}},\ and\ \citenamefont
  {{Huang}}}]{Arvanitaki2015}%
  \BibitemOpen
  \bibfield  {author} {\bibinfo {author} {\bibfnamefont {A.}~\bibnamefont
  {{Arvanitaki}}}, \bibinfo {author} {\bibfnamefont {M.}~\bibnamefont
  {{Baryakhtar}}}, \ and\ \bibinfo {author} {\bibfnamefont {X.}~\bibnamefont
  {{Huang}}},\ }\href {\doibase 10.1103/PhysRevD.91.084011} {\bibfield
  {journal} {\bibinfo  {journal} {\prd}\ }\textbf {\bibinfo {volume} {91}},\
  \bibinfo {eid} {084011} (\bibinfo {year} {2015})},\ \Eprint
  {http://arxiv.org/abs/1411.2263} {arXiv:1411.2263 [hep-ph]} \BibitemShut
  {NoStop}%
\bibitem [{\citenamefont {Hook}\ \emph
  {et~al.}(2018{\natexlab{a}})\citenamefont {Hook}, \citenamefont {Kahn},
  \citenamefont {Safdi},\ and\ \citenamefont {Sun}}]{hook2018}%
  \BibitemOpen
  \bibfield  {author} {\bibinfo {author} {\bibfnamefont {A.}~\bibnamefont
  {Hook}}, \bibinfo {author} {\bibfnamefont {Y.}~\bibnamefont {Kahn}}, \bibinfo
  {author} {\bibfnamefont {B.}~\bibnamefont {Safdi}}, \ and\ \bibinfo {author}
  {\bibfnamefont {Z.}~\bibnamefont {Sun}},\ }\href@noop {} {\bibfield
  {journal} {\bibinfo  {journal} {Phys.\ Rev.\ Lett.}\ }\textbf {\bibinfo
  {volume} {121}},\ \bibinfo {pages} {241102} (\bibinfo {year}
  {2018}{\natexlab{a}})}\BibitemShut {NoStop}%
\bibitem [{\citenamefont {Huang}\ \emph {et~al.}(2018)\citenamefont {Huang},
  \citenamefont {Kadota}, \citenamefont {Sekiguchi},\ and\ \citenamefont
  {Tashiro}}]{huang2018}%
  \BibitemOpen
  \bibfield  {author} {\bibinfo {author} {\bibfnamefont {F.~P.}\ \bibnamefont
  {Huang}}, \bibinfo {author} {\bibfnamefont {K.}~\bibnamefont {Kadota}},
  \bibinfo {author} {\bibfnamefont {T.}~\bibnamefont {Sekiguchi}}, \ and\
  \bibinfo {author} {\bibfnamefont {H.}~\bibnamefont {Tashiro}},\ }\href
  {\doibase 10.1103/PhysRevD.97.123001} {\bibfield  {journal} {\bibinfo
  {journal} {Physical Review D}\ }\textbf {\bibinfo {volume} {97}},\ \bibinfo
  {pages} {123001} (\bibinfo {year} {2018})}\BibitemShut {NoStop}%
\bibitem [{\citenamefont {{Mukherjee}}\ \emph {et~al.}(2018)\citenamefont
  {{Mukherjee}}, \citenamefont {{Khatri}},\ and\ \citenamefont
  {{Wandelt}}}]{mukherjee2018}%
  \BibitemOpen
  \bibfield  {author} {\bibinfo {author} {\bibfnamefont {S.}~\bibnamefont
  {{Mukherjee}}}, \bibinfo {author} {\bibfnamefont {R.}~\bibnamefont
  {{Khatri}}}, \ and\ \bibinfo {author} {\bibfnamefont {B.~D.}\ \bibnamefont
  {{Wandelt}}},\ }\href {\doibase 10.1088/1475-7516/2018/04/045} {\bibfield
  {journal} {\bibinfo  {journal} {JCAP}\ }\textbf {\bibinfo {volume} {2018}},\
  \bibinfo {eid} {045} (\bibinfo {year} {2018})},\ \Eprint
  {http://arxiv.org/abs/1801.09701} {arXiv:1801.09701 [astro-ph.CO]}
  \BibitemShut {NoStop}%
\bibitem [{\citenamefont {{Day}}\ and\ \citenamefont
  {{McDonald}}(2019)}]{day2019}%
  \BibitemOpen
  \bibfield  {author} {\bibinfo {author} {\bibfnamefont {F.~V.}\ \bibnamefont
  {{Day}}}\ and\ \bibinfo {author} {\bibfnamefont {J.~I.}\ \bibnamefont
  {{McDonald}}},\ }\href {\doibase 10.1088/1475-7516/2019/10/051} {\bibfield
  {journal} {\bibinfo  {journal} {JCAP}\ }\textbf {\bibinfo {volume} {2019}},\
  \bibinfo {eid} {051} (\bibinfo {year} {2019})},\ \Eprint
  {http://arxiv.org/abs/1904.08341} {arXiv:1904.08341 [hep-ph]} \BibitemShut
  {NoStop}%
\bibitem [{\citenamefont {{Battye}}\ \emph {et~al.}(2020)\citenamefont
  {{Battye}}, \citenamefont {{Garbrecht}}, \citenamefont {{McDonald}},
  \citenamefont {{Pace}},\ and\ \citenamefont {{Srinivasan}}}]{battye2020}%
  \BibitemOpen
  \bibfield  {author} {\bibinfo {author} {\bibfnamefont {R.~A.}\ \bibnamefont
  {{Battye}}}, \bibinfo {author} {\bibfnamefont {B.}~\bibnamefont
  {{Garbrecht}}}, \bibinfo {author} {\bibfnamefont {J.~I.}\ \bibnamefont
  {{McDonald}}}, \bibinfo {author} {\bibfnamefont {F.}~\bibnamefont {{Pace}}},
  \ and\ \bibinfo {author} {\bibfnamefont {S.}~\bibnamefont {{Srinivasan}}},\
  }\href {\doibase 10.1103/PhysRevD.102.023504} {\bibfield  {journal} {\bibinfo
   {journal} {\prd}\ }\textbf {\bibinfo {volume} {102}},\ \bibinfo {eid}
  {023504} (\bibinfo {year} {2020})},\ \Eprint
  {http://arxiv.org/abs/1910.11907} {arXiv:1910.11907 [astro-ph.CO]}
  \BibitemShut {NoStop}%
\bibitem [{\citenamefont {{Leroy}}\ \emph {et~al.}(2020)\citenamefont
  {{Leroy}}, \citenamefont {{Chianese}}, \citenamefont {{Edwards}},\ and\
  \citenamefont {{Weniger}}}]{leroy2020}%
  \BibitemOpen
  \bibfield  {author} {\bibinfo {author} {\bibfnamefont {M.}~\bibnamefont
  {{Leroy}}}, \bibinfo {author} {\bibfnamefont {M.}~\bibnamefont {{Chianese}}},
  \bibinfo {author} {\bibfnamefont {T.~D.~P.}\ \bibnamefont {{Edwards}}}, \
  and\ \bibinfo {author} {\bibfnamefont {C.}~\bibnamefont {{Weniger}}},\ }\href
  {\doibase 10.1103/PhysRevD.101.123003} {\bibfield  {journal} {\bibinfo
  {journal} {\prd}\ }\textbf {\bibinfo {volume} {101}},\ \bibinfo {eid}
  {123003} (\bibinfo {year} {2020})},\ \Eprint
  {http://arxiv.org/abs/1912.08815} {arXiv:1912.08815 [hep-ph]} \BibitemShut
  {NoStop}%
\bibitem [{\citenamefont {{Edwards}}\ \emph {et~al.}(2020)\citenamefont
  {{Edwards}}, \citenamefont {{Chianese}}, \citenamefont {{Kavanagh}},
  \citenamefont {{Nissanke}},\ and\ \citenamefont {{Weniger}}}]{edwards2020}%
  \BibitemOpen
  \bibfield  {author} {\bibinfo {author} {\bibfnamefont {T.~D.~P.}\
  \bibnamefont {{Edwards}}}, \bibinfo {author} {\bibfnamefont {M.}~\bibnamefont
  {{Chianese}}}, \bibinfo {author} {\bibfnamefont {B.~J.}\ \bibnamefont
  {{Kavanagh}}}, \bibinfo {author} {\bibfnamefont {S.~M.}\ \bibnamefont
  {{Nissanke}}}, \ and\ \bibinfo {author} {\bibfnamefont {C.}~\bibnamefont
  {{Weniger}}},\ }\href {\doibase 10.1103/PhysRevLett.124.161101} {\bibfield
  {journal} {\bibinfo  {journal} {\prl}\ }\textbf {\bibinfo {volume} {124}},\
  \bibinfo {eid} {161101} (\bibinfo {year} {2020})},\ \Eprint
  {http://arxiv.org/abs/1905.04686} {arXiv:1905.04686 [hep-ph]} \BibitemShut
  {NoStop}%
\bibitem [{\citenamefont {{Mukherjee}}\ \emph {et~al.}(2020)\citenamefont
  {{Mukherjee}}, \citenamefont {{Spergel}}, \citenamefont {{Khatri}},\ and\
  \citenamefont {{Wand elt}}}]{mukherjee2020}%
  \BibitemOpen
  \bibfield  {author} {\bibinfo {author} {\bibfnamefont {S.}~\bibnamefont
  {{Mukherjee}}}, \bibinfo {author} {\bibfnamefont {D.~N.}\ \bibnamefont
  {{Spergel}}}, \bibinfo {author} {\bibfnamefont {R.}~\bibnamefont {{Khatri}}},
  \ and\ \bibinfo {author} {\bibfnamefont {B.~D.}\ \bibnamefont {{Wand elt}}},\
  }\href {\doibase 10.1088/1475-7516/2020/02/032} {\bibfield  {journal}
  {\bibinfo  {journal} {JCAP}\ }\textbf {\bibinfo {volume} {2020}},\ \bibinfo
  {eid} {032} (\bibinfo {year} {2020})},\ \Eprint
  {http://arxiv.org/abs/1908.07534} {arXiv:1908.07534 [astro-ph.CO]}
  \BibitemShut {NoStop}%
\bibitem [{\citenamefont {Mori}\ and\ \citenamefont {et~al.}(2013)}]{mori2013}%
  \BibitemOpen
  \bibfield  {author} {\bibinfo {author} {\bibfnamefont {K.}~\bibnamefont
  {Mori}}\ and\ \bibinfo {author} {\bibnamefont {et~al.}},\ }\href@noop {}
  {\bibfield  {journal} {\bibinfo  {journal} {ApJ}\ }\textbf {\bibinfo {volume}
  {770}},\ \bibinfo {pages} {L23} (\bibinfo {year} {2013})}\BibitemShut
  {NoStop}%
\bibitem [{Note1()}]{Note1}%
  \BibitemOpen
  \bibinfo {note} {The National Radio Astronomy Observatory is a facility of
  the National Science Foundation operated under cooperative agreement by
  Associated Universities, Inc.}\BibitemShut {Stop}%
\bibitem [{Note2()}]{Note2}%
  \BibitemOpen
  \bibinfo {note} {McMullin, J. P., Waters, B., Schiebel, D., Young, W., \&
  Golap, K. 2007, Astronomical Data Analysis Software and Systems XVI (ASP
  Conf. Ser. 376), ed. R. A. Shaw, F. Hill, \& D. J. Bell (San Francisco, CA:
  ASP), 127. See also \protect \url
  {https://science.nrao.edu/facilities/vla/data-processing/pipeline} for the
  CASA calibration pipeline and \protect \url
  {https://science.nrao.edu/facilities/vla/data-processing} for VLA calibration
  and analysis.}\BibitemShut {Stop}%
\bibitem [{\citenamefont {Bower}\ \emph {et~al.}(2015)\citenamefont {Bower},
  \citenamefont {Deller}, \citenamefont {Demorest}, \citenamefont {Brunthaler},
  \citenamefont {Falcke}, \citenamefont {Moscibrodzka}, \citenamefont
  {O'Leary}, \citenamefont {Eatough}, \citenamefont {Kramer}, \citenamefont
  {Lee}, \citenamefont {Spitler}, \citenamefont {Desvignes}, \citenamefont
  {Rushton}, \citenamefont {Doeleman},\ and\ \citenamefont {Reid}}]{Bower2015}%
  \BibitemOpen
  \bibfield  {author} {\bibinfo {author} {\bibfnamefont {G.~C.}\ \bibnamefont
  {Bower}}, \bibinfo {author} {\bibfnamefont {A.}~\bibnamefont {Deller}},
  \bibinfo {author} {\bibfnamefont {P.}~\bibnamefont {Demorest}}, \bibinfo
  {author} {\bibfnamefont {A.}~\bibnamefont {Brunthaler}}, \bibinfo {author}
  {\bibfnamefont {H.}~\bibnamefont {Falcke}}, \bibinfo {author} {\bibfnamefont
  {M.}~\bibnamefont {Moscibrodzka}}, \bibinfo {author} {\bibfnamefont {R.~M.}\
  \bibnamefont {O'Leary}}, \bibinfo {author} {\bibfnamefont {R.~P.}\
  \bibnamefont {Eatough}}, \bibinfo {author} {\bibfnamefont {M.}~\bibnamefont
  {Kramer}}, \bibinfo {author} {\bibfnamefont {K.~J.}\ \bibnamefont {Lee}},
  \bibinfo {author} {\bibfnamefont {L.}~\bibnamefont {Spitler}}, \bibinfo
  {author} {\bibfnamefont {G.}~\bibnamefont {Desvignes}}, \bibinfo {author}
  {\bibfnamefont {A.~P.}\ \bibnamefont {Rushton}}, \bibinfo {author}
  {\bibfnamefont {S.}~\bibnamefont {Doeleman}}, \ and\ \bibinfo {author}
  {\bibfnamefont {M.~J.}\ \bibnamefont {Reid}},\ }\href {\doibase
  10.1088/0004-637X/798/2/120} {\bibfield  {journal} {\bibinfo  {journal} {The
  Astrophysical Journal}\ }\textbf {\bibinfo {volume} {798}},\ \bibinfo {pages}
  {120} (\bibinfo {year} {2015})}\BibitemShut {NoStop}%
\bibitem [{\citenamefont {Kennea}\ and\ \citenamefont
  {et~al.}(2013)}]{kennea2013}%
  \BibitemOpen
  \bibfield  {author} {\bibinfo {author} {\bibfnamefont {J.~A.}\ \bibnamefont
  {Kennea}}\ and\ \bibinfo {author} {\bibnamefont {et~al.}},\ }\href@noop {}
  {\bibfield  {journal} {\bibinfo  {journal} {ApJ}\ }\textbf {\bibinfo {volume}
  {770}},\ \bibinfo {pages} {L24} (\bibinfo {year} {2013})}\BibitemShut
  {NoStop}%
\bibitem [{\citenamefont {{Price-Whelan}}\ \emph {et~al.}(2018)\citenamefont
  {{Price-Whelan}}, \citenamefont {{Sip{\H{o}}cz}}, \citenamefont
  {{G{\"u}nther}}, \citenamefont {{Lim}}, \citenamefont {{Crawford}},
  \citenamefont {{Conseil}}, \citenamefont {{Shupe}}, \citenamefont {{Craig}},
  \citenamefont {{Dencheva}}, \citenamefont {{Ginsburg}}, \citenamefont
  {{VanderPlas}}, \citenamefont {{Bradley}}, \citenamefont
  {{P{\'e}rez-Su{\'a}rez}}, \citenamefont {{de Val-Borro}}, \citenamefont
  {{Paper Contributors}}, \citenamefont {{Aldcroft}}, \citenamefont {{Cruz}},
  \citenamefont {{Robitaille}}, \citenamefont {{Tollerud}}, \citenamefont
  {{Coordination Committee}}, \citenamefont {{Ardelean}}, \citenamefont
  {{Babej}}, \citenamefont {{Bach}}, \citenamefont {{Bachetti}}, \citenamefont
  {{Bakanov}}, \citenamefont {{Bamford}}, \citenamefont {{Barentsen}},
  \citenamefont {{Barmby}}, \citenamefont {{Baumbach}}, \citenamefont
  {{Berry}}, \citenamefont {{Biscani}}, \citenamefont {{Boquien}},
  \citenamefont {{Bostroem}}, \citenamefont {{Bouma}}, \citenamefont
  {{Brammer}}, \citenamefont {{Bray}}, \citenamefont {{Breytenbach}},
  \citenamefont {{Buddelmeijer}}, \citenamefont {{Burke}}, \citenamefont
  {{Calderone}}, \citenamefont {{Cano Rodr{\'\i}guez}}, \citenamefont {{Cara}},
  \citenamefont {{Cardoso}}, \citenamefont {{Cheedella}}, \citenamefont
  {{Copin}}, \citenamefont {{Corrales}}, \citenamefont {{Crichton}},
  \citenamefont {{D{\textquoteright}Avella}}, \citenamefont {{Deil}},
  \citenamefont {{Depagne}}, \citenamefont {{Dietrich}}, \citenamefont
  {{Donath}}, \citenamefont {{Droettboom}}, \citenamefont {{Earl}},
  \citenamefont {{Erben}}, \citenamefont {{Fabbro}}, \citenamefont
  {{Ferreira}}, \citenamefont {{Finethy}}, \citenamefont {{Fox}}, \citenamefont
  {{Garrison}}, \citenamefont {{Gibbons}}, \citenamefont {{Goldstein}},
  \citenamefont {{Gommers}}, \citenamefont {{Greco}}, \citenamefont
  {{Greenfield}}, \citenamefont {{Groener}}, \citenamefont {{Grollier}},
  \citenamefont {{Hagen}}, \citenamefont {{Hirst}}, \citenamefont {{Homeier}},
  \citenamefont {{Horton}}, \citenamefont {{Hosseinzadeh}}, \citenamefont
  {{Hu}}, \citenamefont {{Hunkeler}}, \citenamefont {{Ivezi{\'c}}},
  \citenamefont {{Jain}}, \citenamefont {{Jenness}}, \citenamefont {{Kanarek}},
  \citenamefont {{Kendrew}}, \citenamefont {{Kern}}, \citenamefont
  {{Kerzendorf}}, \citenamefont {{Khvalko}}, \citenamefont {{King}},
  \citenamefont {{Kirkby}}, \citenamefont {{Kulkarni}}, \citenamefont
  {{Kumar}}, \citenamefont {{Lee}}, \citenamefont {{Lenz}}, \citenamefont
  {{Littlefair}}, \citenamefont {{Ma}}, \citenamefont {{Macleod}},
  \citenamefont {{Mastropietro}}, \citenamefont {{McCully}}, \citenamefont
  {{Montagnac}}, \citenamefont {{Morris}}, \citenamefont {{Mueller}},
  \citenamefont {{Mumford}}, \citenamefont {{Muna}}, \citenamefont {{Murphy}},
  \citenamefont {{Nelson}}, \citenamefont {{Nguyen}}, \citenamefont {{Ninan}},
  \citenamefont {{N{\"o}the}}, \citenamefont {{Ogaz}}, \citenamefont {{Oh}},
  \citenamefont {{Parejko}}, \citenamefont {{Parley}}, \citenamefont
  {{Pascual}}, \citenamefont {{Patil}}, \citenamefont {{Patil}}, \citenamefont
  {{Plunkett}}, \citenamefont {{Prochaska}}, \citenamefont {{Rastogi}},
  \citenamefont {{Reddy Janga}}, \citenamefont {{Sabater}}, \citenamefont
  {{Sakurikar}}, \citenamefont {{Seifert}}, \citenamefont {{Sherbert}},
  \citenamefont {{Sherwood-Taylor}}, \citenamefont {{Shih}}, \citenamefont
  {{Sick}}, \citenamefont {{Silbiger}}, \citenamefont {{Singanamalla}},
  \citenamefont {{Singer}}, \citenamefont {{Sladen}}, \citenamefont {{Sooley}},
  \citenamefont {{Sornarajah}}, \citenamefont {{Streicher}}, \citenamefont
  {{Teuben}}, \citenamefont {{Thomas}}, \citenamefont {{Tremblay}},
  \citenamefont {{Turner}}, \citenamefont {{Terr{\'o}n}}, \citenamefont {{van
  Kerkwijk}}, \citenamefont {{de la Vega}}, \citenamefont {{Watkins}},
  \citenamefont {{Weaver}}, \citenamefont {{Whitmore}}, \citenamefont
  {{Woillez}}, \citenamefont {{Zabalza}},\ and\ \citenamefont
  {{Contributors}}}]{astropy:2018}%
  \BibitemOpen
  \bibfield  {author} {\bibinfo {author} {\bibfnamefont {A.~M.}\ \bibnamefont
  {{Price-Whelan}}}, \bibinfo {author} {\bibfnamefont {B.~M.}\ \bibnamefont
  {{Sip{\H{o}}cz}}}, \bibinfo {author} {\bibfnamefont {H.~M.}\ \bibnamefont
  {{G{\"u}nther}}}, \bibinfo {author} {\bibfnamefont {P.~L.}\ \bibnamefont
  {{Lim}}}, \bibinfo {author} {\bibfnamefont {S.~M.}\ \bibnamefont
  {{Crawford}}}, \bibinfo {author} {\bibfnamefont {S.}~\bibnamefont
  {{Conseil}}}, \bibinfo {author} {\bibfnamefont {D.~L.}\ \bibnamefont
  {{Shupe}}}, \bibinfo {author} {\bibfnamefont {M.~W.}\ \bibnamefont
  {{Craig}}}, \bibinfo {author} {\bibfnamefont {N.}~\bibnamefont {{Dencheva}}},
  \bibinfo {author} {\bibfnamefont {A.}~\bibnamefont {{Ginsburg}}}, \bibinfo
  {author} {\bibfnamefont {J.~T.}\ \bibnamefont {{VanderPlas}}}, \bibinfo
  {author} {\bibfnamefont {L.~D.}\ \bibnamefont {{Bradley}}}, \bibinfo {author}
  {\bibfnamefont {D.}~\bibnamefont {{P{\'e}rez-Su{\'a}rez}}}, \bibinfo {author}
  {\bibfnamefont {M.}~\bibnamefont {{de Val-Borro}}}, \bibinfo {author}
  {\bibfnamefont {P.}~\bibnamefont {{Paper Contributors}}}, \bibinfo {author}
  {\bibfnamefont {T.~L.}\ \bibnamefont {{Aldcroft}}}, \bibinfo {author}
  {\bibfnamefont {K.~L.}\ \bibnamefont {{Cruz}}}, \bibinfo {author}
  {\bibfnamefont {T.~P.}\ \bibnamefont {{Robitaille}}}, \bibinfo {author}
  {\bibfnamefont {E.~J.}\ \bibnamefont {{Tollerud}}}, \bibinfo {author}
  {\bibfnamefont {A.}~\bibnamefont {{Coordination Committee}}}, \bibinfo
  {author} {\bibfnamefont {C.}~\bibnamefont {{Ardelean}}}, \bibinfo {author}
  {\bibfnamefont {T.}~\bibnamefont {{Babej}}}, \bibinfo {author} {\bibfnamefont
  {Y.~P.}\ \bibnamefont {{Bach}}}, \bibinfo {author} {\bibfnamefont
  {M.}~\bibnamefont {{Bachetti}}}, \bibinfo {author} {\bibfnamefont {A.~V.}\
  \bibnamefont {{Bakanov}}}, \bibinfo {author} {\bibfnamefont {S.~P.}\
  \bibnamefont {{Bamford}}}, \bibinfo {author} {\bibfnamefont {G.}~\bibnamefont
  {{Barentsen}}}, \bibinfo {author} {\bibfnamefont {P.}~\bibnamefont
  {{Barmby}}}, \bibinfo {author} {\bibfnamefont {A.}~\bibnamefont
  {{Baumbach}}}, \bibinfo {author} {\bibfnamefont {K.~L.}\ \bibnamefont
  {{Berry}}}, \bibinfo {author} {\bibfnamefont {F.}~\bibnamefont {{Biscani}}},
  \bibinfo {author} {\bibfnamefont {M.}~\bibnamefont {{Boquien}}}, \bibinfo
  {author} {\bibfnamefont {K.~A.}\ \bibnamefont {{Bostroem}}}, \bibinfo
  {author} {\bibfnamefont {L.~G.}\ \bibnamefont {{Bouma}}}, \bibinfo {author}
  {\bibfnamefont {G.~B.}\ \bibnamefont {{Brammer}}}, \bibinfo {author}
  {\bibfnamefont {E.~M.}\ \bibnamefont {{Bray}}}, \bibinfo {author}
  {\bibfnamefont {H.}~\bibnamefont {{Breytenbach}}}, \bibinfo {author}
  {\bibfnamefont {H.}~\bibnamefont {{Buddelmeijer}}}, \bibinfo {author}
  {\bibfnamefont {D.~J.}\ \bibnamefont {{Burke}}}, \bibinfo {author}
  {\bibfnamefont {G.}~\bibnamefont {{Calderone}}}, \bibinfo {author}
  {\bibfnamefont {J.~L.}\ \bibnamefont {{Cano Rodr{\'\i}guez}}}, \bibinfo
  {author} {\bibfnamefont {M.}~\bibnamefont {{Cara}}}, \bibinfo {author}
  {\bibfnamefont {J.~V.~M.}\ \bibnamefont {{Cardoso}}}, \bibinfo {author}
  {\bibfnamefont {S.}~\bibnamefont {{Cheedella}}}, \bibinfo {author}
  {\bibfnamefont {Y.}~\bibnamefont {{Copin}}}, \bibinfo {author} {\bibfnamefont
  {L.}~\bibnamefont {{Corrales}}}, \bibinfo {author} {\bibfnamefont
  {D.}~\bibnamefont {{Crichton}}}, \bibinfo {author} {\bibfnamefont
  {D.}~\bibnamefont {{D{\textquoteright}Avella}}}, \bibinfo {author}
  {\bibfnamefont {C.}~\bibnamefont {{Deil}}}, \bibinfo {author} {\bibfnamefont
  {{\'E}.}~\bibnamefont {{Depagne}}}, \bibinfo {author} {\bibfnamefont {J.~P.}\
  \bibnamefont {{Dietrich}}}, \bibinfo {author} {\bibfnamefont
  {A.}~\bibnamefont {{Donath}}}, \bibinfo {author} {\bibfnamefont
  {M.}~\bibnamefont {{Droettboom}}}, \bibinfo {author} {\bibfnamefont
  {N.}~\bibnamefont {{Earl}}}, \bibinfo {author} {\bibfnamefont
  {T.}~\bibnamefont {{Erben}}}, \bibinfo {author} {\bibfnamefont
  {S.}~\bibnamefont {{Fabbro}}}, \bibinfo {author} {\bibfnamefont {L.~A.}\
  \bibnamefont {{Ferreira}}}, \bibinfo {author} {\bibfnamefont
  {T.}~\bibnamefont {{Finethy}}}, \bibinfo {author} {\bibfnamefont {R.~T.}\
  \bibnamefont {{Fox}}}, \bibinfo {author} {\bibfnamefont {L.~H.}\ \bibnamefont
  {{Garrison}}}, \bibinfo {author} {\bibfnamefont {S.~L.~J.}\ \bibnamefont
  {{Gibbons}}}, \bibinfo {author} {\bibfnamefont {D.~A.}\ \bibnamefont
  {{Goldstein}}}, \bibinfo {author} {\bibfnamefont {R.}~\bibnamefont
  {{Gommers}}}, \bibinfo {author} {\bibfnamefont {J.~P.}\ \bibnamefont
  {{Greco}}}, \bibinfo {author} {\bibfnamefont {P.}~\bibnamefont
  {{Greenfield}}}, \bibinfo {author} {\bibfnamefont {A.~M.}\ \bibnamefont
  {{Groener}}}, \bibinfo {author} {\bibfnamefont {F.}~\bibnamefont
  {{Grollier}}}, \bibinfo {author} {\bibfnamefont {A.}~\bibnamefont {{Hagen}}},
  \bibinfo {author} {\bibfnamefont {P.}~\bibnamefont {{Hirst}}}, \bibinfo
  {author} {\bibfnamefont {D.}~\bibnamefont {{Homeier}}}, \bibinfo {author}
  {\bibfnamefont {A.~J.}\ \bibnamefont {{Horton}}}, \bibinfo {author}
  {\bibfnamefont {G.}~\bibnamefont {{Hosseinzadeh}}}, \bibinfo {author}
  {\bibfnamefont {L.}~\bibnamefont {{Hu}}}, \bibinfo {author} {\bibfnamefont
  {J.~S.}\ \bibnamefont {{Hunkeler}}}, \bibinfo {author} {\bibfnamefont
  {{\v{Z}}.}~\bibnamefont {{Ivezi{\'c}}}}, \bibinfo {author} {\bibfnamefont
  {A.}~\bibnamefont {{Jain}}}, \bibinfo {author} {\bibfnamefont
  {T.}~\bibnamefont {{Jenness}}}, \bibinfo {author} {\bibfnamefont
  {G.}~\bibnamefont {{Kanarek}}}, \bibinfo {author} {\bibfnamefont
  {S.}~\bibnamefont {{Kendrew}}}, \bibinfo {author} {\bibfnamefont {N.~S.}\
  \bibnamefont {{Kern}}}, \bibinfo {author} {\bibfnamefont {W.~E.}\
  \bibnamefont {{Kerzendorf}}}, \bibinfo {author} {\bibfnamefont
  {A.}~\bibnamefont {{Khvalko}}}, \bibinfo {author} {\bibfnamefont
  {J.}~\bibnamefont {{King}}}, \bibinfo {author} {\bibfnamefont
  {D.}~\bibnamefont {{Kirkby}}}, \bibinfo {author} {\bibfnamefont {A.~M.}\
  \bibnamefont {{Kulkarni}}}, \bibinfo {author} {\bibfnamefont
  {A.}~\bibnamefont {{Kumar}}}, \bibinfo {author} {\bibfnamefont
  {A.}~\bibnamefont {{Lee}}}, \bibinfo {author} {\bibfnamefont
  {D.}~\bibnamefont {{Lenz}}}, \bibinfo {author} {\bibfnamefont {S.~P.}\
  \bibnamefont {{Littlefair}}}, \bibinfo {author} {\bibfnamefont
  {Z.}~\bibnamefont {{Ma}}}, \bibinfo {author} {\bibfnamefont {D.~M.}\
  \bibnamefont {{Macleod}}}, \bibinfo {author} {\bibfnamefont {M.}~\bibnamefont
  {{Mastropietro}}}, \bibinfo {author} {\bibfnamefont {C.}~\bibnamefont
  {{McCully}}}, \bibinfo {author} {\bibfnamefont {S.}~\bibnamefont
  {{Montagnac}}}, \bibinfo {author} {\bibfnamefont {B.~M.}\ \bibnamefont
  {{Morris}}}, \bibinfo {author} {\bibfnamefont {M.}~\bibnamefont {{Mueller}}},
  \bibinfo {author} {\bibfnamefont {S.~J.}\ \bibnamefont {{Mumford}}}, \bibinfo
  {author} {\bibfnamefont {D.}~\bibnamefont {{Muna}}}, \bibinfo {author}
  {\bibfnamefont {N.~A.}\ \bibnamefont {{Murphy}}}, \bibinfo {author}
  {\bibfnamefont {S.}~\bibnamefont {{Nelson}}}, \bibinfo {author}
  {\bibfnamefont {G.~H.}\ \bibnamefont {{Nguyen}}}, \bibinfo {author}
  {\bibfnamefont {J.~P.}\ \bibnamefont {{Ninan}}}, \bibinfo {author}
  {\bibfnamefont {M.}~\bibnamefont {{N{\"o}the}}}, \bibinfo {author}
  {\bibfnamefont {S.}~\bibnamefont {{Ogaz}}}, \bibinfo {author} {\bibfnamefont
  {S.}~\bibnamefont {{Oh}}}, \bibinfo {author} {\bibfnamefont {J.~K.}\
  \bibnamefont {{Parejko}}}, \bibinfo {author} {\bibfnamefont {N.}~\bibnamefont
  {{Parley}}}, \bibinfo {author} {\bibfnamefont {S.}~\bibnamefont {{Pascual}}},
  \bibinfo {author} {\bibfnamefont {R.}~\bibnamefont {{Patil}}}, \bibinfo
  {author} {\bibfnamefont {A.~A.}\ \bibnamefont {{Patil}}}, \bibinfo {author}
  {\bibfnamefont {A.~L.}\ \bibnamefont {{Plunkett}}}, \bibinfo {author}
  {\bibfnamefont {J.~X.}\ \bibnamefont {{Prochaska}}}, \bibinfo {author}
  {\bibfnamefont {T.}~\bibnamefont {{Rastogi}}}, \bibinfo {author}
  {\bibfnamefont {V.}~\bibnamefont {{Reddy Janga}}}, \bibinfo {author}
  {\bibfnamefont {J.}~\bibnamefont {{Sabater}}}, \bibinfo {author}
  {\bibfnamefont {P.}~\bibnamefont {{Sakurikar}}}, \bibinfo {author}
  {\bibfnamefont {M.}~\bibnamefont {{Seifert}}}, \bibinfo {author}
  {\bibfnamefont {L.~E.}\ \bibnamefont {{Sherbert}}}, \bibinfo {author}
  {\bibfnamefont {H.}~\bibnamefont {{Sherwood-Taylor}}}, \bibinfo {author}
  {\bibfnamefont {A.~Y.}\ \bibnamefont {{Shih}}}, \bibinfo {author}
  {\bibfnamefont {J.}~\bibnamefont {{Sick}}}, \bibinfo {author} {\bibfnamefont
  {M.~T.}\ \bibnamefont {{Silbiger}}}, \bibinfo {author} {\bibfnamefont
  {S.}~\bibnamefont {{Singanamalla}}}, \bibinfo {author} {\bibfnamefont
  {L.~P.}\ \bibnamefont {{Singer}}}, \bibinfo {author} {\bibfnamefont {P.~H.}\
  \bibnamefont {{Sladen}}}, \bibinfo {author} {\bibfnamefont {K.~A.}\
  \bibnamefont {{Sooley}}}, \bibinfo {author} {\bibfnamefont {S.}~\bibnamefont
  {{Sornarajah}}}, \bibinfo {author} {\bibfnamefont {O.}~\bibnamefont
  {{Streicher}}}, \bibinfo {author} {\bibfnamefont {P.}~\bibnamefont
  {{Teuben}}}, \bibinfo {author} {\bibfnamefont {S.~W.}\ \bibnamefont
  {{Thomas}}}, \bibinfo {author} {\bibfnamefont {G.~R.}\ \bibnamefont
  {{Tremblay}}}, \bibinfo {author} {\bibfnamefont {J.~E.~H.}\ \bibnamefont
  {{Turner}}}, \bibinfo {author} {\bibfnamefont {V.}~\bibnamefont
  {{Terr{\'o}n}}}, \bibinfo {author} {\bibfnamefont {M.~H.}\ \bibnamefont {{van
  Kerkwijk}}}, \bibinfo {author} {\bibfnamefont {A.}~\bibnamefont {{de la
  Vega}}}, \bibinfo {author} {\bibfnamefont {L.~L.}\ \bibnamefont {{Watkins}}},
  \bibinfo {author} {\bibfnamefont {B.~A.}\ \bibnamefont {{Weaver}}}, \bibinfo
  {author} {\bibfnamefont {J.~B.}\ \bibnamefont {{Whitmore}}}, \bibinfo
  {author} {\bibfnamefont {J.}~\bibnamefont {{Woillez}}}, \bibinfo {author}
  {\bibfnamefont {V.}~\bibnamefont {{Zabalza}}}, \ and\ \bibinfo {author}
  {\bibfnamefont {A.}~\bibnamefont {{Contributors}}},\ }\href {\doibase
  10.3847/1538-3881/aabc4f} {\bibfield  {journal} {\bibinfo  {journal} {AJ}\
  }\textbf {\bibinfo {volume} {156}},\ \bibinfo {eid} {123} (\bibinfo {year}
  {2018})}\BibitemShut {NoStop}%
\bibitem [{Note3()}]{Note3}%
  \BibitemOpen
  \bibinfo {note} {See Supplemental Material at [URL].}\BibitemShut {Stop}%
\bibitem [{Note4()}]{Note4}%
  \BibitemOpen
  \bibinfo {note} {Our assumption about the expected signal bandwidth does
  depend on models for the magnetar, but is $\protect \mathcal
  {O}(v_0/c)$.}\BibitemShut {Stop}%
\bibitem [{\citenamefont {Goldreich}\ and\ \citenamefont
  {Julian}(1969)}]{goldreich1969}%
  \BibitemOpen
  \bibfield  {author} {\bibinfo {author} {\bibfnamefont {P.}~\bibnamefont
  {Goldreich}}\ and\ \bibinfo {author} {\bibfnamefont {W.~H.}\ \bibnamefont
  {Julian}},\ }\href@noop {} {\bibfield  {journal} {\bibinfo  {journal} {ApJ}\
  }\textbf {\bibinfo {volume} {157}},\ \bibinfo {pages} {869} (\bibinfo {year}
  {1969})}\BibitemShut {NoStop}%
\bibitem [{\citenamefont {{Navarro}}\ \emph {et~al.}(1996)\citenamefont
  {{Navarro}}, \citenamefont {{Frenk}},\ and\ \citenamefont {{White}}}]{NFW}%
  \BibitemOpen
  \bibfield  {author} {\bibinfo {author} {\bibfnamefont {J.~F.}\ \bibnamefont
  {{Navarro}}}, \bibinfo {author} {\bibfnamefont {C.~S.}\ \bibnamefont
  {{Frenk}}}, \ and\ \bibinfo {author} {\bibfnamefont {S.~D.~M.}\ \bibnamefont
  {{White}}},\ }\href {\doibase 10.1086/177173} {\bibfield  {journal} {\bibinfo
   {journal} {ApJ}\ }\textbf {\bibinfo {volume} {462}},\ \bibinfo {pages} {563}
  (\bibinfo {year} {1996})},\ \Eprint {http://arxiv.org/abs/astro-ph/9508025}
  {arXiv:astro-ph/9508025 [astro-ph]} \BibitemShut {NoStop}%
\bibitem [{\citenamefont {{McMillan}}(2017)}]{McMillan2017}%
  \BibitemOpen
  \bibfield  {author} {\bibinfo {author} {\bibfnamefont {P.~J.}\ \bibnamefont
  {{McMillan}}},\ }\href {\doibase 10.1093/mnras/stw2759} {\bibfield  {journal}
  {\bibinfo  {journal} {MNRAS}\ }\textbf {\bibinfo {volume} {465}},\ \bibinfo
  {pages} {76} (\bibinfo {year} {2017})},\ \Eprint
  {http://arxiv.org/abs/1608.00971} {arXiv:1608.00971 [astro-ph.GA]}
  \BibitemShut {NoStop}%
\bibitem [{\citenamefont {Abuter}\ \emph {et~al.}(2019)\citenamefont {Abuter},
  \citenamefont {Amorim}, \citenamefont {Baub{\"{o}}ck}, \citenamefont
  {Berger}, \citenamefont {Bonnet}, \citenamefont {Brandner}, \citenamefont
  {Cl{\'{e}}net}, \citenamefont {{Coud{\'{e}} du Foresto}}, \citenamefont
  {de~Zeeuw}, \citenamefont {Dexter}, \citenamefont {Duvert}, \citenamefont
  {Eckart}, \citenamefont {Eisenhauer}, \citenamefont {{F{\"{o}}rster
  Schreiber}}, \citenamefont {Garcia}, \citenamefont {Gao}, \citenamefont
  {Gendron}, \citenamefont {Genzel}, \citenamefont {Gerhard}, \citenamefont
  {Gillessen}, \citenamefont {Habibi}, \citenamefont {Haubois}, \citenamefont
  {Henning}, \citenamefont {Hippler}, \citenamefont {Horrobin}, \citenamefont
  {Jim{\'{e}}nez-Rosales}, \citenamefont {Jocou}, \citenamefont {Kervella},
  \citenamefont {Lacour}, \citenamefont {Lapeyr{\`{e}}re}, \citenamefont {{Le
  Bouquin}}, \citenamefont {L{\'{e}}na}, \citenamefont {Ott}, \citenamefont
  {Paumard}, \citenamefont {Perraut}, \citenamefont {Perrin}, \citenamefont
  {Pfuhl}, \citenamefont {Rabien}, \citenamefont {{Rodriguez Coira}},
  \citenamefont {Rousset}, \citenamefont {Scheithauer}, \citenamefont
  {Sternberg}, \citenamefont {Straub}, \citenamefont {Straubmeier},
  \citenamefont {Sturm}, \citenamefont {Tacconi}, \citenamefont {Vincent},
  \citenamefont {von Fellenberg}, \citenamefont {Waisberg}, \citenamefont
  {Widmann}, \citenamefont {Wieprecht}, \citenamefont {Wiezorrek},
  \citenamefont {Woillez},\ and\ \citenamefont {Yazici}}]{Abuter2019}%
  \BibitemOpen
  \bibfield  {author} {\bibinfo {author} {\bibfnamefont {R.}~\bibnamefont
  {Abuter}}, \bibinfo {author} {\bibfnamefont {A.}~\bibnamefont {Amorim}},
  \bibinfo {author} {\bibfnamefont {M.}~\bibnamefont {Baub{\"{o}}ck}}, \bibinfo
  {author} {\bibfnamefont {J.~P.}\ \bibnamefont {Berger}}, \bibinfo {author}
  {\bibfnamefont {H.}~\bibnamefont {Bonnet}}, \bibinfo {author} {\bibfnamefont
  {W.}~\bibnamefont {Brandner}}, \bibinfo {author} {\bibfnamefont
  {Y.}~\bibnamefont {Cl{\'{e}}net}}, \bibinfo {author} {\bibfnamefont
  {V.}~\bibnamefont {{Coud{\'{e}} du Foresto}}}, \bibinfo {author}
  {\bibfnamefont {P.~T.}\ \bibnamefont {de~Zeeuw}}, \bibinfo {author}
  {\bibfnamefont {J.}~\bibnamefont {Dexter}}, \bibinfo {author} {\bibfnamefont
  {G.}~\bibnamefont {Duvert}}, \bibinfo {author} {\bibfnamefont
  {A.}~\bibnamefont {Eckart}}, \bibinfo {author} {\bibfnamefont
  {F.}~\bibnamefont {Eisenhauer}}, \bibinfo {author} {\bibfnamefont {N.~M.}\
  \bibnamefont {{F{\"{o}}rster Schreiber}}}, \bibinfo {author} {\bibfnamefont
  {P.}~\bibnamefont {Garcia}}, \bibinfo {author} {\bibfnamefont
  {F.}~\bibnamefont {Gao}}, \bibinfo {author} {\bibfnamefont {E.}~\bibnamefont
  {Gendron}}, \bibinfo {author} {\bibfnamefont {R.}~\bibnamefont {Genzel}},
  \bibinfo {author} {\bibfnamefont {O.}~\bibnamefont {Gerhard}}, \bibinfo
  {author} {\bibfnamefont {S.}~\bibnamefont {Gillessen}}, \bibinfo {author}
  {\bibfnamefont {M.}~\bibnamefont {Habibi}}, \bibinfo {author} {\bibfnamefont
  {X.}~\bibnamefont {Haubois}}, \bibinfo {author} {\bibfnamefont
  {T.}~\bibnamefont {Henning}}, \bibinfo {author} {\bibfnamefont
  {S.}~\bibnamefont {Hippler}}, \bibinfo {author} {\bibfnamefont
  {M.}~\bibnamefont {Horrobin}}, \bibinfo {author} {\bibfnamefont
  {A.}~\bibnamefont {Jim{\'{e}}nez-Rosales}}, \bibinfo {author} {\bibfnamefont
  {L.}~\bibnamefont {Jocou}}, \bibinfo {author} {\bibfnamefont
  {P.}~\bibnamefont {Kervella}}, \bibinfo {author} {\bibfnamefont
  {S.}~\bibnamefont {Lacour}}, \bibinfo {author} {\bibfnamefont
  {V.}~\bibnamefont {Lapeyr{\`{e}}re}}, \bibinfo {author} {\bibfnamefont
  {J.-B.}\ \bibnamefont {{Le Bouquin}}}, \bibinfo {author} {\bibfnamefont
  {P.}~\bibnamefont {L{\'{e}}na}}, \bibinfo {author} {\bibfnamefont
  {T.}~\bibnamefont {Ott}}, \bibinfo {author} {\bibfnamefont {T.}~\bibnamefont
  {Paumard}}, \bibinfo {author} {\bibfnamefont {K.}~\bibnamefont {Perraut}},
  \bibinfo {author} {\bibfnamefont {G.}~\bibnamefont {Perrin}}, \bibinfo
  {author} {\bibfnamefont {O.}~\bibnamefont {Pfuhl}}, \bibinfo {author}
  {\bibfnamefont {S.}~\bibnamefont {Rabien}}, \bibinfo {author} {\bibfnamefont
  {G.}~\bibnamefont {{Rodriguez Coira}}}, \bibinfo {author} {\bibfnamefont
  {G.}~\bibnamefont {Rousset}}, \bibinfo {author} {\bibfnamefont
  {S.}~\bibnamefont {Scheithauer}}, \bibinfo {author} {\bibfnamefont
  {A.}~\bibnamefont {Sternberg}}, \bibinfo {author} {\bibfnamefont
  {O.}~\bibnamefont {Straub}}, \bibinfo {author} {\bibfnamefont
  {C.}~\bibnamefont {Straubmeier}}, \bibinfo {author} {\bibfnamefont
  {E.}~\bibnamefont {Sturm}}, \bibinfo {author} {\bibfnamefont {L.~J.}\
  \bibnamefont {Tacconi}}, \bibinfo {author} {\bibfnamefont {F.}~\bibnamefont
  {Vincent}}, \bibinfo {author} {\bibfnamefont {S.}~\bibnamefont {von
  Fellenberg}}, \bibinfo {author} {\bibfnamefont {I.}~\bibnamefont {Waisberg}},
  \bibinfo {author} {\bibfnamefont {F.}~\bibnamefont {Widmann}}, \bibinfo
  {author} {\bibfnamefont {E.}~\bibnamefont {Wieprecht}}, \bibinfo {author}
  {\bibfnamefont {E.}~\bibnamefont {Wiezorrek}}, \bibinfo {author}
  {\bibfnamefont {J.}~\bibnamefont {Woillez}}, \ and\ \bibinfo {author}
  {\bibfnamefont {S.}~\bibnamefont {Yazici}},\ }\href {\doibase
  10.1051/0004-6361/201935656} {\bibfield  {journal} {\bibinfo  {journal}
  {Astronomy {\&} Astrophysics}\ }\textbf {\bibinfo {volume} {625}},\ \bibinfo
  {pages} {L10} (\bibinfo {year} {2019})}\BibitemShut {NoStop}%
\bibitem [{\citenamefont {Lacroix}(2018)}]{Lacroix2018}%
  \BibitemOpen
  \bibfield  {author} {\bibinfo {author} {\bibfnamefont {T.}~\bibnamefont
  {Lacroix}},\ }\href {\doibase 10.1051/0004-6361/201832652} {\bibfield
  {journal} {\bibinfo  {journal} {Astronomy and Astrophysics}\ }\textbf
  {\bibinfo {volume} {619}},\ \bibinfo {pages} {46} (\bibinfo {year} {2018})},\
  \Eprint {http://arxiv.org/abs/1801.01308} {arXiv:1801.01308} \BibitemShut
  {NoStop}%
\bibitem [{\citenamefont {{Di Luzio}}\ \emph {et~al.}(2017)\citenamefont {{Di
  Luzio}}, \citenamefont {Mescia},\ and\ \citenamefont {Nardi}}]{Luzio2017}%
  \BibitemOpen
  \bibfield  {author} {\bibinfo {author} {\bibfnamefont {L.}~\bibnamefont {{Di
  Luzio}}}, \bibinfo {author} {\bibfnamefont {F.}~\bibnamefont {Mescia}}, \
  and\ \bibinfo {author} {\bibfnamefont {E.}~\bibnamefont {Nardi}},\ }\href
  {\doibase 10.1103/PhysRevLett.118.031801} {\bibfield  {journal} {\bibinfo
  {journal} {Physical Review Letters}\ }\textbf {\bibinfo {volume} {118}}
  (\bibinfo {year} {2017}),\ 10.1103/PhysRevLett.118.031801},\ \Eprint
  {http://arxiv.org/abs/1610.07593} {arXiv:1610.07593} \BibitemShut {NoStop}%
\bibitem [{\citenamefont {Anastassopoulos}\ \emph {et~al.}(2017)\citenamefont
  {Anastassopoulos}, \citenamefont {Aune}, \citenamefont {Barth}, \citenamefont
  {Belov}, \citenamefont {Br{\"{a}}uninger}, \citenamefont {Cantatore},
  \citenamefont {Carmona}, \citenamefont {Castel}, \citenamefont {Cetin},
  \citenamefont {Christensen}, \citenamefont {Collar}, \citenamefont {Dafni},
  \citenamefont {Davenport}, \citenamefont {Decker}, \citenamefont {Dermenev},
  \citenamefont {Desch}, \citenamefont {Eleftheriadis}, \citenamefont
  {Fanourakis}, \citenamefont {Ferrer-Ribas}, \citenamefont {Fischer},
  \citenamefont {Garc{\'{i}}a}, \citenamefont {Gardikiotis}, \citenamefont
  {Garza}, \citenamefont {Gazis}, \citenamefont {Geralis}, \citenamefont
  {Giomataris}, \citenamefont {Gninenko}, \citenamefont {Hailey}, \citenamefont
  {Hasinoff}, \citenamefont {Hoffmann}, \citenamefont {Iguaz}, \citenamefont
  {Irastorza}, \citenamefont {Jakobsen}, \citenamefont {Jacoby}, \citenamefont
  {Jakovcic}, \citenamefont {Kaminski}, \citenamefont {Karuza}, \citenamefont
  {Kralj}, \citenamefont {Krcmar}, \citenamefont {Kostoglou}, \citenamefont
  {Krieger}, \citenamefont {Lakic}, \citenamefont {Laurent}, \citenamefont
  {Liolios}, \citenamefont {Ljubicic}, \citenamefont {Luz{\'{o}}n},
  \citenamefont {Maroudas}, \citenamefont {Miceli}, \citenamefont {Neff},
  \citenamefont {Ortega}, \citenamefont {Papaevangelou}, \citenamefont
  {Paraschou}, \citenamefont {Pivovaroff}, \citenamefont {Raffelt},
  \citenamefont {Rosu}, \citenamefont {Ruz}, \citenamefont {Ch{\'{o}}liz},
  \citenamefont {Savvidis}, \citenamefont {Schmidt}, \citenamefont
  {Semertzidis}, \citenamefont {Solanki}, \citenamefont {Stewart},
  \citenamefont {Vafeiadis}, \citenamefont {Vogel}, \citenamefont {Yildiz},\
  and\ \citenamefont {Zioutas}}]{Anastassopoulos2017}%
  \BibitemOpen
  \bibfield  {author} {\bibinfo {author} {\bibfnamefont {V.}~\bibnamefont
  {Anastassopoulos}}, \bibinfo {author} {\bibfnamefont {S.}~\bibnamefont
  {Aune}}, \bibinfo {author} {\bibfnamefont {K.}~\bibnamefont {Barth}},
  \bibinfo {author} {\bibfnamefont {A.}~\bibnamefont {Belov}}, \bibinfo
  {author} {\bibfnamefont {H.}~\bibnamefont {Br{\"{a}}uninger}}, \bibinfo
  {author} {\bibfnamefont {G.}~\bibnamefont {Cantatore}}, \bibinfo {author}
  {\bibfnamefont {J.~M.}\ \bibnamefont {Carmona}}, \bibinfo {author}
  {\bibfnamefont {J.~F.}\ \bibnamefont {Castel}}, \bibinfo {author}
  {\bibfnamefont {S.~A.}\ \bibnamefont {Cetin}}, \bibinfo {author}
  {\bibfnamefont {F.}~\bibnamefont {Christensen}}, \bibinfo {author}
  {\bibfnamefont {J.~I.}\ \bibnamefont {Collar}}, \bibinfo {author}
  {\bibfnamefont {T.}~\bibnamefont {Dafni}}, \bibinfo {author} {\bibfnamefont
  {M.}~\bibnamefont {Davenport}}, \bibinfo {author} {\bibfnamefont {T.~A.}\
  \bibnamefont {Decker}}, \bibinfo {author} {\bibfnamefont {A.}~\bibnamefont
  {Dermenev}}, \bibinfo {author} {\bibfnamefont {K.}~\bibnamefont {Desch}},
  \bibinfo {author} {\bibfnamefont {C.}~\bibnamefont {Eleftheriadis}}, \bibinfo
  {author} {\bibfnamefont {G.}~\bibnamefont {Fanourakis}}, \bibinfo {author}
  {\bibfnamefont {E.}~\bibnamefont {Ferrer-Ribas}}, \bibinfo {author}
  {\bibfnamefont {H.}~\bibnamefont {Fischer}}, \bibinfo {author} {\bibfnamefont
  {J.~A.}\ \bibnamefont {Garc{\'{i}}a}}, \bibinfo {author} {\bibfnamefont
  {A.}~\bibnamefont {Gardikiotis}}, \bibinfo {author} {\bibfnamefont {J.~G.}\
  \bibnamefont {Garza}}, \bibinfo {author} {\bibfnamefont {E.~N.}\ \bibnamefont
  {Gazis}}, \bibinfo {author} {\bibfnamefont {T.}~\bibnamefont {Geralis}},
  \bibinfo {author} {\bibfnamefont {I.}~\bibnamefont {Giomataris}}, \bibinfo
  {author} {\bibfnamefont {S.}~\bibnamefont {Gninenko}}, \bibinfo {author}
  {\bibfnamefont {C.~J.}\ \bibnamefont {Hailey}}, \bibinfo {author}
  {\bibfnamefont {M.~D.}\ \bibnamefont {Hasinoff}}, \bibinfo {author}
  {\bibfnamefont {D.~H.}\ \bibnamefont {Hoffmann}}, \bibinfo {author}
  {\bibfnamefont {F.~J.}\ \bibnamefont {Iguaz}}, \bibinfo {author}
  {\bibfnamefont {I.~G.}\ \bibnamefont {Irastorza}}, \bibinfo {author}
  {\bibfnamefont {A.}~\bibnamefont {Jakobsen}}, \bibinfo {author}
  {\bibfnamefont {J.}~\bibnamefont {Jacoby}}, \bibinfo {author} {\bibfnamefont
  {K.}~\bibnamefont {Jakovcic}}, \bibinfo {author} {\bibfnamefont
  {J.}~\bibnamefont {Kaminski}}, \bibinfo {author} {\bibfnamefont
  {M.}~\bibnamefont {Karuza}}, \bibinfo {author} {\bibfnamefont
  {N.}~\bibnamefont {Kralj}}, \bibinfo {author} {\bibfnamefont
  {M.}~\bibnamefont {Krcmar}}, \bibinfo {author} {\bibfnamefont
  {S.}~\bibnamefont {Kostoglou}}, \bibinfo {author} {\bibfnamefont
  {C.}~\bibnamefont {Krieger}}, \bibinfo {author} {\bibfnamefont
  {B.}~\bibnamefont {Lakic}}, \bibinfo {author} {\bibfnamefont {J.~M.}\
  \bibnamefont {Laurent}}, \bibinfo {author} {\bibfnamefont {A.}~\bibnamefont
  {Liolios}}, \bibinfo {author} {\bibfnamefont {A.}~\bibnamefont {Ljubicic}},
  \bibinfo {author} {\bibfnamefont {G.}~\bibnamefont {Luz{\'{o}}n}}, \bibinfo
  {author} {\bibfnamefont {M.}~\bibnamefont {Maroudas}}, \bibinfo {author}
  {\bibfnamefont {L.}~\bibnamefont {Miceli}}, \bibinfo {author} {\bibfnamefont
  {S.}~\bibnamefont {Neff}}, \bibinfo {author} {\bibfnamefont {I.}~\bibnamefont
  {Ortega}}, \bibinfo {author} {\bibfnamefont {T.}~\bibnamefont
  {Papaevangelou}}, \bibinfo {author} {\bibfnamefont {K.}~\bibnamefont
  {Paraschou}}, \bibinfo {author} {\bibfnamefont {M.~J.}\ \bibnamefont
  {Pivovaroff}}, \bibinfo {author} {\bibfnamefont {G.}~\bibnamefont {Raffelt}},
  \bibinfo {author} {\bibfnamefont {M.}~\bibnamefont {Rosu}}, \bibinfo {author}
  {\bibfnamefont {J.}~\bibnamefont {Ruz}}, \bibinfo {author} {\bibfnamefont
  {E.~R.}\ \bibnamefont {Ch{\'{o}}liz}}, \bibinfo {author} {\bibfnamefont
  {I.}~\bibnamefont {Savvidis}}, \bibinfo {author} {\bibfnamefont
  {S.}~\bibnamefont {Schmidt}}, \bibinfo {author} {\bibfnamefont {Y.~K.}\
  \bibnamefont {Semertzidis}}, \bibinfo {author} {\bibfnamefont {S.~K.}\
  \bibnamefont {Solanki}}, \bibinfo {author} {\bibfnamefont {L.}~\bibnamefont
  {Stewart}}, \bibinfo {author} {\bibfnamefont {T.}~\bibnamefont {Vafeiadis}},
  \bibinfo {author} {\bibfnamefont {J.~K.}\ \bibnamefont {Vogel}}, \bibinfo
  {author} {\bibfnamefont {S.~C.}\ \bibnamefont {Yildiz}}, \ and\ \bibinfo
  {author} {\bibfnamefont {K.}~\bibnamefont {Zioutas}},\ }\href {\doibase
  10.1038/nphys4109} {\bibfield  {journal} {\bibinfo  {journal} {Nature
  Physics}\ }\textbf {\bibinfo {volume} {13}},\ \bibinfo {pages} {584}
  (\bibinfo {year} {2017})},\ \Eprint {http://arxiv.org/abs/1705.02290}
  {arXiv:1705.02290} \BibitemShut {NoStop}%
\bibitem [{\citenamefont {Caputo}\ \emph {et~al.}(2019)\citenamefont {Caputo},
  \citenamefont {Regis}, \citenamefont {Taoso},\ and\ \citenamefont
  {Witte}}]{Caputo2019}%
  \BibitemOpen
  \bibfield  {author} {\bibinfo {author} {\bibfnamefont {A.}~\bibnamefont
  {Caputo}}, \bibinfo {author} {\bibfnamefont {M.}~\bibnamefont {Regis}},
  \bibinfo {author} {\bibfnamefont {M.}~\bibnamefont {Taoso}}, \ and\ \bibinfo
  {author} {\bibfnamefont {S.~J.}\ \bibnamefont {Witte}},\ }\href {\doibase
  10.1088/1475-7516/2019/03/027} {\bibfield  {journal} {\bibinfo  {journal}
  {Journal of Cosmology and Astroparticle Physics}\ }\textbf {\bibinfo {volume}
  {2019}} (\bibinfo {year} {2019}),\ 10.1088/1475-7516/2019/03/027},\ \Eprint
  {http://arxiv.org/abs/1811.08436} {arXiv:1811.08436} \BibitemShut {NoStop}%
\bibitem [{\citenamefont {{Cautun}}\ \emph {et~al.}(2020)\citenamefont
  {{Cautun}}, \citenamefont {{Ben{\'\i}tez-Llambay}}, \citenamefont {{Deason}},
  \citenamefont {{Frenk}}, \citenamefont {{Fattahi}}, \citenamefont
  {{G{\'o}mez}}, \citenamefont {{Grand}}, \citenamefont {{Oman}}, \citenamefont
  {{Navarro}},\ and\ \citenamefont {{Simpson}}}]{cautun2020}%
  \BibitemOpen
  \bibfield  {author} {\bibinfo {author} {\bibfnamefont {M.}~\bibnamefont
  {{Cautun}}}, \bibinfo {author} {\bibfnamefont {A.}~\bibnamefont
  {{Ben{\'\i}tez-Llambay}}}, \bibinfo {author} {\bibfnamefont {A.~J.}\
  \bibnamefont {{Deason}}}, \bibinfo {author} {\bibfnamefont {C.~S.}\
  \bibnamefont {{Frenk}}}, \bibinfo {author} {\bibfnamefont {A.}~\bibnamefont
  {{Fattahi}}}, \bibinfo {author} {\bibfnamefont {F.~A.}\ \bibnamefont
  {{G{\'o}mez}}}, \bibinfo {author} {\bibfnamefont {R.~J.~J.}\ \bibnamefont
  {{Grand}}}, \bibinfo {author} {\bibfnamefont {K.~A.}\ \bibnamefont {{Oman}}},
  \bibinfo {author} {\bibfnamefont {J.~F.}\ \bibnamefont {{Navarro}}}, \ and\
  \bibinfo {author} {\bibfnamefont {C.~M.}\ \bibnamefont {{Simpson}}},\ }\href
  {\doibase 10.1093/mnras/staa1017} {\bibfield  {journal} {\bibinfo  {journal}
  {MNRAS}\ }\textbf {\bibinfo {volume} {494}},\ \bibinfo {pages} {4291}
  (\bibinfo {year} {2020})},\ \Eprint {http://arxiv.org/abs/1911.04557}
  {arXiv:1911.04557 [astro-ph.GA]} \BibitemShut {NoStop}%
\bibitem [{\citenamefont {Hooper}(2017)}]{hooper2017}%
  \BibitemOpen
  \bibfield  {author} {\bibinfo {author} {\bibfnamefont {D.}~\bibnamefont
  {Hooper}},\ }\href {\doibase 10.1016/j.dark.2016.11.005} {\bibfield
  {journal} {\bibinfo  {journal} {Physics of the Dark Universe}\ }\textbf
  {\bibinfo {volume} {15}},\ \bibinfo {pages} {53} (\bibinfo {year} {2017})},\
  \Eprint {http://arxiv.org/abs/1608.00003} {arXiv:1608.00003} \BibitemShut
  {NoStop}%
\bibitem [{\citenamefont {Safdi}\ \emph {et~al.}(2019)\citenamefont {Safdi},
  \citenamefont {Sun},\ and\ \citenamefont {Chen}}]{Safdi2019}%
  \BibitemOpen
  \bibfield  {author} {\bibinfo {author} {\bibfnamefont {B.~R.}\ \bibnamefont
  {Safdi}}, \bibinfo {author} {\bibfnamefont {Z.}~\bibnamefont {Sun}}, \ and\
  \bibinfo {author} {\bibfnamefont {A.~Y.}\ \bibnamefont {Chen}},\ }\href
  {\doibase 10.1103/PhysRevD.99.123021} {\bibfield  {journal} {\bibinfo
  {journal} {Physical Review D}\ }\textbf {\bibinfo {volume} {99}} (\bibinfo
  {year} {2019}),\ 10.1103/PhysRevD.99.123021},\ \Eprint
  {http://arxiv.org/abs/1811.01020} {arXiv:1811.01020} \BibitemShut {NoStop}%
\bibitem [{\citenamefont {{McMullin}}\ \emph {et~al.}(2007)\citenamefont
  {{McMullin}}, \citenamefont {{Waters}}, \citenamefont {{Schiebel}},
  \citenamefont {{Young}},\ and\ \citenamefont {{Golap}}}]{CASA}%
  \BibitemOpen
  \bibfield  {author} {\bibinfo {author} {\bibfnamefont {J.~P.}\ \bibnamefont
  {{McMullin}}}, \bibinfo {author} {\bibfnamefont {B.}~\bibnamefont
  {{Waters}}}, \bibinfo {author} {\bibfnamefont {D.}~\bibnamefont
  {{Schiebel}}}, \bibinfo {author} {\bibfnamefont {W.}~\bibnamefont {{Young}}},
  \ and\ \bibinfo {author} {\bibfnamefont {K.}~\bibnamefont {{Golap}}},\ }in\
  \href@noop {} {\emph {\bibinfo {booktitle} {Astronomical Data Analysis
  Software and Systems XVI}}},\ \bibinfo {series} {Astronomical Society of the
  Pacific Conference Series}, Vol.\ \bibinfo {volume} {376},\ \bibinfo {editor}
  {edited by\ \bibinfo {editor} {\bibfnamefont {R.~A.}\ \bibnamefont {{Shaw}}},
  \bibinfo {editor} {\bibfnamefont {F.}~\bibnamefont {{Hill}}}, \ and\ \bibinfo
  {editor} {\bibfnamefont {D.~J.}\ \bibnamefont {{Bell}}}}\ (\bibinfo {year}
  {2007})\ p.\ \bibinfo {pages} {127}\BibitemShut {NoStop}%
\bibitem [{\citenamefont {{van der Walt}}\ \emph {et~al.}(2011)\citenamefont
  {{van der Walt}}, \citenamefont {{Colbert}},\ and\ \citenamefont
  {{Varoquaux}}}]{NumPy}%
  \BibitemOpen
  \bibfield  {author} {\bibinfo {author} {\bibfnamefont {S.}~\bibnamefont {{van
  der Walt}}}, \bibinfo {author} {\bibfnamefont {S.~C.}\ \bibnamefont
  {{Colbert}}}, \ and\ \bibinfo {author} {\bibfnamefont {G.}~\bibnamefont
  {{Varoquaux}}},\ }\href@noop {} {\bibfield  {journal} {\bibinfo  {journal}
  {Computing in Science Engineering}\ }\textbf {\bibinfo {volume} {13}},\
  \bibinfo {pages} {22} (\bibinfo {year} {2011})}\BibitemShut {NoStop}%
\bibitem [{\citenamefont {{Hunter}}(2007)}]{Matplotlib}%
  \BibitemOpen
  \bibfield  {author} {\bibinfo {author} {\bibfnamefont {J.~D.}\ \bibnamefont
  {{Hunter}}},\ }\href@noop {} {\bibfield  {journal} {\bibinfo  {journal}
  {Computing in Science Engineering}\ }\textbf {\bibinfo {volume} {9}},\
  \bibinfo {pages} {90} (\bibinfo {year} {2007})}\BibitemShut {NoStop}%
\bibitem [{Note5()}]{Note5}%
  \BibitemOpen
  \bibinfo {note} {Http://www.astropy.org}\BibitemShut {NoStop}%
\bibitem [{\citenamefont {{Astropy Collaboration}}\ \emph
  {et~al.}(2013)\citenamefont {{Astropy Collaboration}}, \citenamefont
  {{Robitaille}}, \citenamefont {{Tollerud}}, \citenamefont {{Greenfield}},
  \citenamefont {{Droettboom}}, \citenamefont {{Bray}}, \citenamefont
  {{Aldcroft}}, \citenamefont {{Davis}}, \citenamefont {{Ginsburg}},
  \citenamefont {{Price-Whelan}}, \citenamefont {{Kerzendorf}}, \citenamefont
  {{Conley}}, \citenamefont {{Crighton}}, \citenamefont {{Barbary}},
  \citenamefont {{Muna}}, \citenamefont {{Ferguson}}, \citenamefont
  {{Grollier}}, \citenamefont {{Parikh}}, \citenamefont {{Nair}}, \citenamefont
  {{Unther}}, \citenamefont {{Deil}}, \citenamefont {{Woillez}}, \citenamefont
  {{Conseil}}, \citenamefont {{Kramer}}, \citenamefont {{Turner}},
  \citenamefont {{Singer}}, \citenamefont {{Fox}}, \citenamefont {{Weaver}},
  \citenamefont {{Zabalza}}, \citenamefont {{Edwards}}, \citenamefont {{Azalee
  Bostroem}}, \citenamefont {{Burke}}, \citenamefont {{Casey}}, \citenamefont
  {{Crawford}}, \citenamefont {{Dencheva}}, \citenamefont {{Ely}},
  \citenamefont {{Jenness}}, \citenamefont {{Labrie}}, \citenamefont {{Lim}},
  \citenamefont {{Pierfederici}}, \citenamefont {{Pontzen}}, \citenamefont
  {{Ptak}}, \citenamefont {{Refsdal}}, \citenamefont {{Servillat}},\ and\
  \citenamefont {{Streicher}}}]{astropy:2013}%
  \BibitemOpen
  \bibfield  {author} {\bibinfo {author} {\bibnamefont {{Astropy
  Collaboration}}}, \bibinfo {author} {\bibfnamefont {T.~P.}\ \bibnamefont
  {{Robitaille}}}, \bibinfo {author} {\bibfnamefont {E.~J.}\ \bibnamefont
  {{Tollerud}}}, \bibinfo {author} {\bibfnamefont {P.}~\bibnamefont
  {{Greenfield}}}, \bibinfo {author} {\bibfnamefont {M.}~\bibnamefont
  {{Droettboom}}}, \bibinfo {author} {\bibfnamefont {E.}~\bibnamefont
  {{Bray}}}, \bibinfo {author} {\bibfnamefont {T.}~\bibnamefont {{Aldcroft}}},
  \bibinfo {author} {\bibfnamefont {M.}~\bibnamefont {{Davis}}}, \bibinfo
  {author} {\bibfnamefont {A.}~\bibnamefont {{Ginsburg}}}, \bibinfo {author}
  {\bibfnamefont {A.~M.}\ \bibnamefont {{Price-Whelan}}}, \bibinfo {author}
  {\bibfnamefont {W.~E.}\ \bibnamefont {{Kerzendorf}}}, \bibinfo {author}
  {\bibfnamefont {A.}~\bibnamefont {{Conley}}}, \bibinfo {author}
  {\bibfnamefont {N.}~\bibnamefont {{Crighton}}}, \bibinfo {author}
  {\bibfnamefont {K.}~\bibnamefont {{Barbary}}}, \bibinfo {author}
  {\bibfnamefont {D.}~\bibnamefont {{Muna}}}, \bibinfo {author} {\bibfnamefont
  {H.}~\bibnamefont {{Ferguson}}}, \bibinfo {author} {\bibfnamefont
  {F.}~\bibnamefont {{Grollier}}}, \bibinfo {author} {\bibfnamefont {M.~M.}\
  \bibnamefont {{Parikh}}}, \bibinfo {author} {\bibfnamefont {P.~H.}\
  \bibnamefont {{Nair}}}, \bibinfo {author} {\bibfnamefont {H.~M.}\
  \bibnamefont {{Unther}}}, \bibinfo {author} {\bibfnamefont {C.}~\bibnamefont
  {{Deil}}}, \bibinfo {author} {\bibfnamefont {J.}~\bibnamefont {{Woillez}}},
  \bibinfo {author} {\bibfnamefont {S.}~\bibnamefont {{Conseil}}}, \bibinfo
  {author} {\bibfnamefont {R.}~\bibnamefont {{Kramer}}}, \bibinfo {author}
  {\bibfnamefont {J.~E.~H.}\ \bibnamefont {{Turner}}}, \bibinfo {author}
  {\bibfnamefont {L.}~\bibnamefont {{Singer}}}, \bibinfo {author}
  {\bibfnamefont {R.}~\bibnamefont {{Fox}}}, \bibinfo {author} {\bibfnamefont
  {B.~A.}\ \bibnamefont {{Weaver}}}, \bibinfo {author} {\bibfnamefont
  {V.}~\bibnamefont {{Zabalza}}}, \bibinfo {author} {\bibfnamefont {Z.~I.}\
  \bibnamefont {{Edwards}}}, \bibinfo {author} {\bibfnamefont {K.}~\bibnamefont
  {{Azalee Bostroem}}}, \bibinfo {author} {\bibfnamefont {D.~J.}\ \bibnamefont
  {{Burke}}}, \bibinfo {author} {\bibfnamefont {A.~R.}\ \bibnamefont
  {{Casey}}}, \bibinfo {author} {\bibfnamefont {S.~M.}\ \bibnamefont
  {{Crawford}}}, \bibinfo {author} {\bibfnamefont {N.}~\bibnamefont
  {{Dencheva}}}, \bibinfo {author} {\bibfnamefont {J.}~\bibnamefont {{Ely}}},
  \bibinfo {author} {\bibfnamefont {T.}~\bibnamefont {{Jenness}}}, \bibinfo
  {author} {\bibfnamefont {K.}~\bibnamefont {{Labrie}}}, \bibinfo {author}
  {\bibfnamefont {P.~L.}\ \bibnamefont {{Lim}}}, \bibinfo {author}
  {\bibfnamefont {F.}~\bibnamefont {{Pierfederici}}}, \bibinfo {author}
  {\bibfnamefont {A.}~\bibnamefont {{Pontzen}}}, \bibinfo {author}
  {\bibfnamefont {A.}~\bibnamefont {{Ptak}}}, \bibinfo {author} {\bibfnamefont
  {B.}~\bibnamefont {{Refsdal}}}, \bibinfo {author} {\bibfnamefont
  {M.}~\bibnamefont {{Servillat}}}, \ and\ \bibinfo {author} {\bibfnamefont
  {O.}~\bibnamefont {{Streicher}}},\ }\href {\doibase
  10.1051/0004-6361/201322068} {\bibfield  {journal} {\bibinfo  {journal}
  {A\&A}\ }\textbf {\bibinfo {volume} {558}},\ \bibinfo {eid} {A33} (\bibinfo
  {year} {2013})},\ \Eprint {http://arxiv.org/abs/1307.6212} {arXiv:1307.6212
  [astro-ph.IM]} \BibitemShut {NoStop}%
\bibitem [{\citenamefont {Hook}\ \emph
  {et~al.}(2018{\natexlab{b}})\citenamefont {Hook}, \citenamefont {Kahn},
  \citenamefont {Safdi},\ and\ \citenamefont {Sun}}]{hook2018supplement}%
  \BibitemOpen
  \bibfield  {author} {\bibinfo {author} {\bibfnamefont {A.}~\bibnamefont
  {Hook}}, \bibinfo {author} {\bibfnamefont {Y.}~\bibnamefont {Kahn}}, \bibinfo
  {author} {\bibfnamefont {B.~R.}\ \bibnamefont {Safdi}}, \ and\ \bibinfo
  {author} {\bibfnamefont {Z.}~\bibnamefont {Sun}},\ }\href {\doibase
  10.1103/PhysRevLett.121.241102} {\bibfield  {journal} {\bibinfo  {journal}
  {Physical Review Letters}\ }\textbf {\bibinfo {volume} {121}} (\bibinfo
  {year} {2018}{\natexlab{b}}),\ 10.1103/PhysRevLett.121.241102},\ \Eprint
  {http://arxiv.org/abs/1804.03145} {arXiv:1804.03145} \BibitemShut {NoStop}%
\end{thebibliography}%

\clearpage

\onecolumngrid
\begin{center}
  \textbf{A Search for Axionic Dark Matter Using the Magnetar PSR J1745$-$2900\\ {\it  Supplementary Material}}\\
  \vspace{10pt}
  {Jeremy Darling}
  \vspace{10pt}
\end{center}

\setcounter{equation}{0}
\setcounter{figure}{0}
\setcounter{section}{0}
\setcounter{page}{1}
\makeatletter

  This Supplementary Material presents the radio spectra, noise spectra, and significance spectra of the individual bands 
  used to derive limits on the axion-photon coupling $g_{a\gamma\gamma}$ versus
  axion mass $m_a$ presented in the main Letter.  We also present a discussion of the impact of the magnetar's rotation and magnetic
  field axes on the translation of flux density limits into limits on $g_{a\gamma\gamma}$.

\onecolumngrid

  \section{Spectra of the Magnetar PSR J1745$-$2900}

\begin{figure*}[h]
\begin{centering}
  \includegraphics[scale=0.48,trim=0 0 0 0,clip=true]{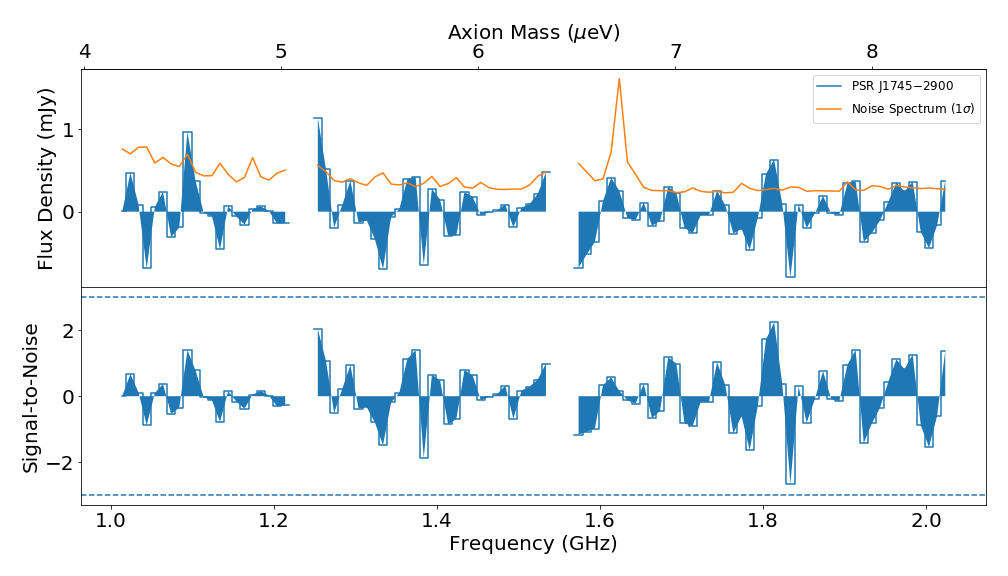}
\caption{
  L-band flux density, noise, and signal-to-noise spectra.
  The upper spectrum provides limits on $g_{a\gamma\gamma}$, while the lower spectrum indicates the
  significance of spectral features.}  \label{fig:Spec_L}
\end{centering}
\end{figure*}

\begin{figure*}
\begin{centering}
  \includegraphics[scale=0.48,trim=0 0 0 0,clip=true]{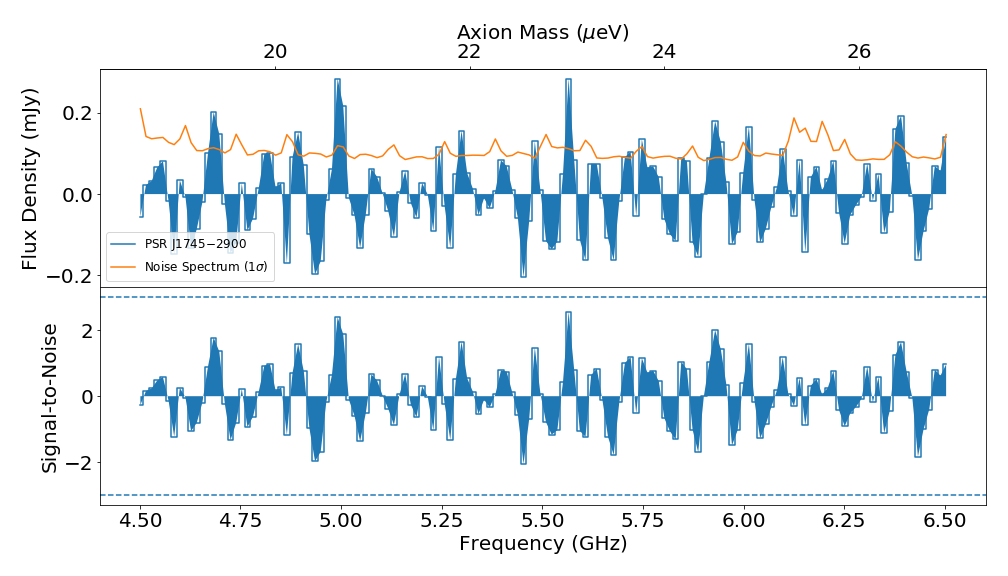}
\caption{
  C-band flux density, noise, and signal-to-noise spectra.}  \label{fig:Spec_C}
\end{centering}
\end{figure*}

\begin{figure*}
\begin{centering}
  \includegraphics[scale=0.48,trim=0 0 0 0,clip=true]{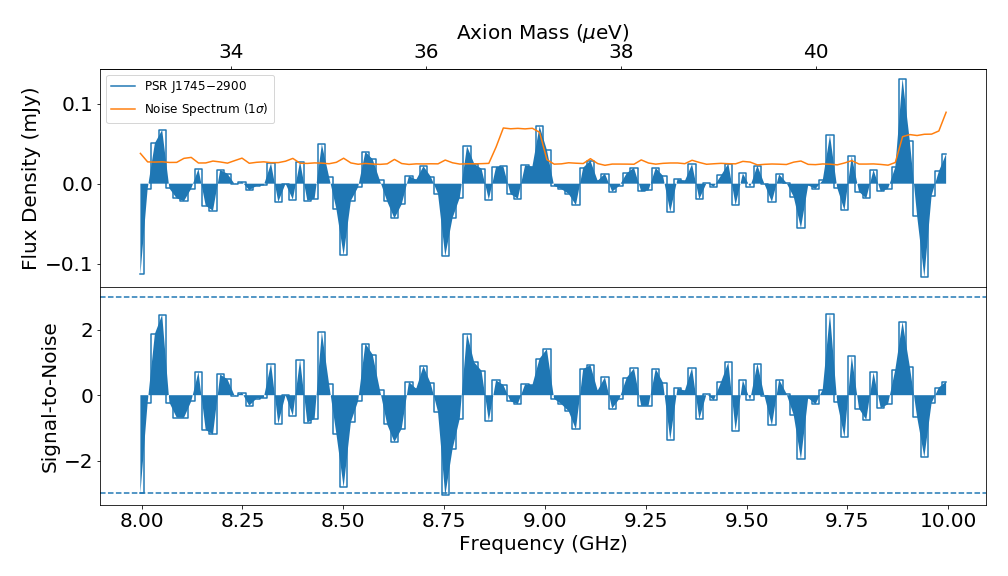}
\caption{
  X-band flux density, noise, and signal-to-noise spectra.}  \label{fig:Spec_X}
\end{centering}
\end{figure*}

\begin{figure*}
\begin{centering}
  \includegraphics[scale=0.48,trim=0 0 0 0,clip=true]{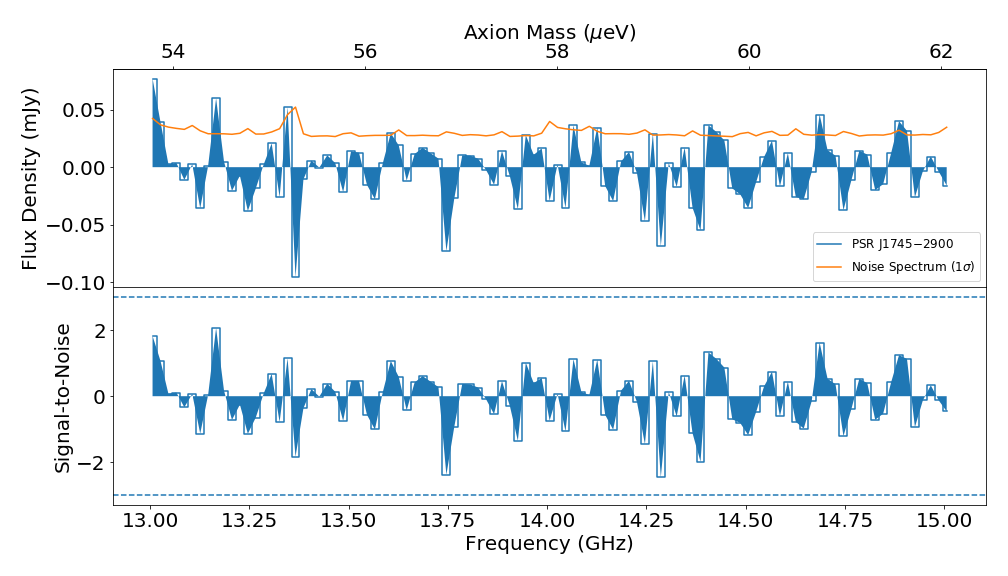}
\caption{
  Ku-band flux density, noise, and signal-to-noise spectra. }
    \label{fig:Spec_Ku}
\end{centering}
\end{figure*}

\begin{figure*}
\begin{centering}
  \includegraphics[scale=0.48,trim=0 0 0 0,clip=true]{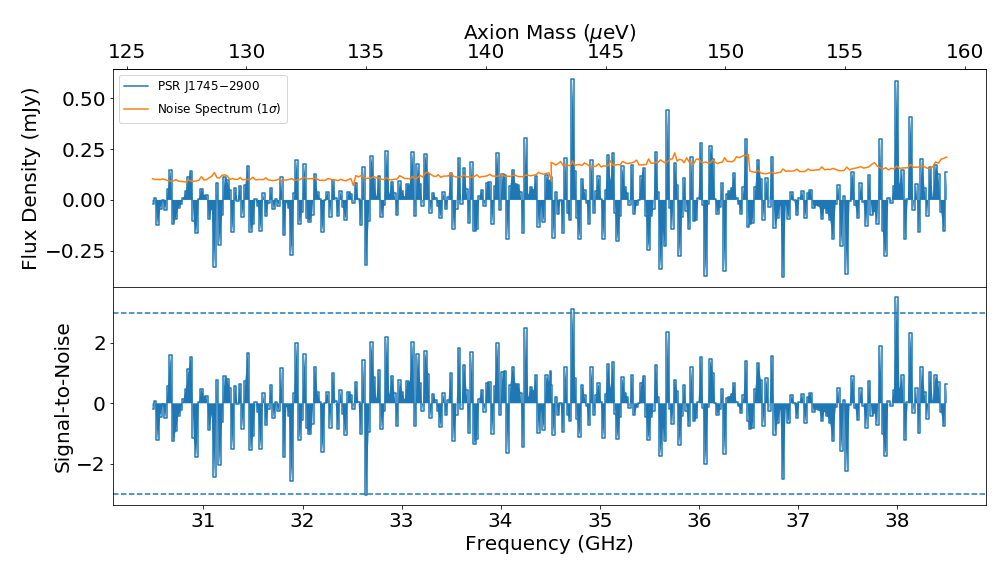}
\caption{
  Ka-band flux density, noise, and signal-to-noise spectra.  The four overlapping basebands
  listed in Table I of the main Letter were combined into a single spectrum.}  \label{fig:Spec_Ka}
\end{centering}
\end{figure*}

Figures \ref{fig:Spec_L}--\ref{fig:Spec_Ka} show the continuum-subtracted flux density spectra, noise spectra, and
significance spectra used for flux density and $g_{a\gamma\gamma}$ limits.

\clearpage

\section{Impact of the Magnetar Spin and Magnetic Field Axes on Axion Limits}

The angular term $(i)$ in Equation 4 of the main Letter modulates the signal, and the modulation time
is less than the integration time of the observations, so we average the expected signal
over the period of the magnetar for each frequency channel.  Moreover, the unknown magnetar
spin axis orientation angle $\theta$ and magnetic dipole offset angle $\theta_m$ impact both the emitted flux
density (main Letter, Equation 3) and the axion-photon conversion radius $r_c$ \cite{hook2018supplement}:
\begin{equation}
  r_c = 224~{\rm km}~\left(r_0 \over 10~{\rm km}\right) \left[ {B_0 \over 10^{14}~{\rm G}} {1\over 2\pi} {\Omega \over 1~{\rm Hz}} \left(4.1~\mu{\rm eV} \over m_a c^2\right)^2\right]^{1/3}  |3\cos\theta\ \hat{m}\cdot\hat{r}-\cos\theta_m|^{1/3}.
\end{equation}
When $r_c < r_0$ no axion conversion occurs, so the rotation period-averaged flux density will be ``censored'' in a
frequency-dependent manner.  The velocity at the conversion point is $v_c^2 \simeq 2G M_{NS}/r_c$ \cite{battye2020}.

\begin{figure}
\begin{centering}
 \includegraphics[scale=0.5,trim=20 20 20 0,clip=true]{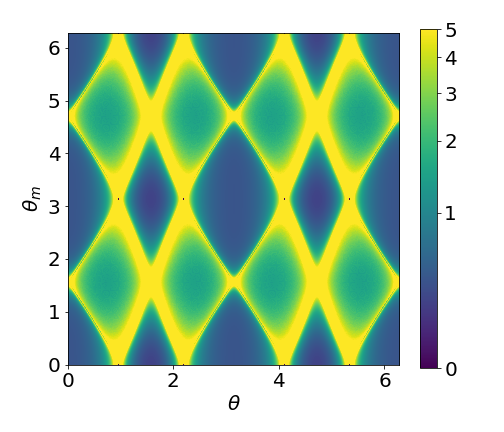}
  \includegraphics[scale=0.52,trim=20 20 20 0,clip=true]{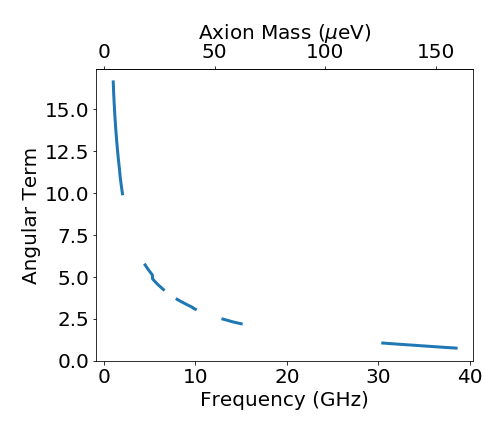}
\caption{
  Left:  Time-averaged angular term at 10~GHz plotted for
the unknown viewing ($\theta$) and magnetic dipole orientation ($\theta_m$) angles of the magnetar.
  Right:  Angle-marginalized and time-averaged angular term $(i)$ versus frequency.
}
\label{fig:angular_term}
\end{centering}
\end{figure}

For each observed frequency channel and possible $(\theta,\theta_m)$ pair, we create an emission
profile over the period of the magnetar rotation that includes no emission when $r_c < r_0$.  We calculate a time-integrated
flux density, and we marginalize these flux densities at each channel over all $(\theta,\theta_m)$.
Figure \ref{fig:angular_term} shows an example at 10~GHz of the influence of $\theta$ and $\theta_m$ on
the angular term $(i)$ that includes time-averaging including times when $r_c < r_0$.  We also plot 
the net result of the marginalization over these unknown angles versus frequency and equivalent axion mass.
These time-averaged and angle-marginalized angular terms are used in the calculation to place limits on
$g_{a\gamma\gamma}$ (main Letter, Equation 4).

\end{document}